\documentclass[11pt
]{article} 
\usepackage[utf8]{inputenc}
\usepackage[a4paper,margin=1in]{geometry}

\usepackage{graphicx}

\usepackage{booktabs} 
\usepackage{array} 
\usepackage{paralist} 
\usepackage{verbatim} 
\usepackage{amsmath}
\usepackage{amssymb}
\usepackage{tikz}
\usetikzlibrary{decorations.pathmorphing}
\usepackage{subcaption}
\usepackage{authblk}
\usepackage{authblk}
\usepackage{microtype}
\usepackage{caption}
\usepackage{datetime}

\renewcommand{\vec}[1]{\boldsymbol{#1}}
\renewcommand{\epsilon}{\varepsilon}
\newcommand{\dd}{\mathrm{d}}
\DeclareMathOperator{\Hankel}{H}

\title{Making Waves Round a Structured Cloak:\\
\emph{Lattices, Negative Refraction and Fringes}}
\author[1]{DJ Colquitt}
\author[2]{IS Jones}
\author[1]{NV Movchan}
\author[1]{AB Movchan}
\author[1,3]{M Brun}
\author[1,4]{RC McPhedran}

\affil[1]{\small Department of Mathematical Sciences, University of Liverpool, Liverpool, L69 3BX, UK}
\affil[2]{School of Engineering, John Moores University, Liverpool, L3 3AF, UK}
\affil[3]{Universit\'{a} di Cagliari, Piazza d’Armi, I-09123 Cagliari, Italy}
\affil[4]{CUDOS, School of Physics, University of Sydney, New South Wales 2006, Australia}

\begin{document}

\maketitle
\begin{abstract}
Using the framework of transformation optics, this paper presents a detailed analysis of a non-singular square cloak for acoustic, out-of-plane shear elastic, and electromagnetic waves.
The generating map is examined in detail and linked to the material properties of the cloak.
Analysis of wave propagation through the cloak is presented and accompanied by numerical illustrations.
The efficacy of the regularised cloak is demonstrated and an objective numerical measure of the quality of the cloaking effect is provided.
It is demonstrated that the cloaking effect persists over a wide range of frequencies.
As a demonstration of the effectiveness of the regularised cloak, a Young's double slit experiment is presented.
The stability of the interference pattern is examined when a cloaked and uncloaked obstacle are successively placed in front of one of the apertures.
This novel link with a well-known quantum mechanical experiment provides an additional method through which the quality of cloaks may be examined.
In the second half of the paper, it is shown that an approximate cloak may be constructed using a discrete lattice structure.
The geometry and material properties of the lattice are derived from the continuum cloak.
The efficiency of the approximate lattice cloak is analysed and a series of illustrative simulations presented.
It is demonstrated that effective cloaking may be obtained by using a relatively simple lattice structure, particularly in the low frequency regime.
\end{abstract}

\begin{center}
\end{center}

\section{Introduction}

Following the publication of two seminal papers in 2006 by Pendry et al.~\cite{Pendry2006} and Leonhardt~\cite{Leonhardt2006}, there has been very considerable interest and many scholarly works have been devoted to the concept of invisibility cloaks; for example, see the recent review article~\cite{Guenneau2011} and references therein.
The experimental validation of cloaking by Schurig et al.~\cite{Schurig2006b} has further increased both scholarly, and popular, interest in invisibility cloaks.
The concept of cloaking via transformation optics is due to an earlier fundamental result by Greenleaf et al.~\cite{Greenleaf2003,Greenleaf2003a} on singular transformations and applications to cloaking for conductivity.
The fundamental property that enables objects to be cloaked from electromagnetic waves is the metric invariance of Maxwell's equations~\cite{Post1962,Ward1996}.

The classical approach to cloaking via transformation geometry involves deforming a region such that a point is mapped to a finite region corresponding to the inner boundary of the cloak.
Usually, such transformations involve transforming a point into a finite region with a smooth boundary, such as an ellipse (see~\cite{Pendry2006,Norris2008,Guenneau2012}, among others).
The mapping is non-singular everywhere, except at the initial point which is deformed into the inner boundary of the cloak.
Since the material properties of the cloak are determined by the transformation~\cite{Schurig2006a}, such singular transformations lead to material properties that are singular.
Greenleaf et al.  addressed this issue in two papers~\cite{Greenleaf2007,Greenleaf2008} and derived the condition for finite energy solutions for weak solutions of the Helmholtz equation and Maxwell's equations.
In~\cite{Greenleaf2007}, it was demonstrated that finite energy solutions to the cloaking problem for the Helmholtz equation exist for an object with a single layer cloak.
However, for the case of Maxwell's equations with internal currents, the cloaking of an infinite cylinder cannot be achieved with a single layer or without imposing a physical surface at the inner boundary of the cloak.
In the same paper, Greenleaf et al. derived an identity linking the transformed scalar wave equation to the metric of the deformed space, which may then by linked to the material properties of the cloak~\cite{Greenleaf2008}.
In 2008 Norris~\cite{Norris2008} studied acoustic cloaking and re-derived an equivalent identity to that in~\cite{Greenleaf2007} using the framework of finite elasticity, leading to a cloak with a density described by a rank two tensor.
Moreover, it was demonstrated that the total mass of the cloak is infinite for the case of perfect cloaking.
Norris further demonstrated that the problem of infinite mass could be overcome if both the density and elastic properties of the cloak were anisotropic.
An alternative approach to negate the problem of singular material properties is to construct a so-called \emph{near cloak} by regularising the transformation~\cite{Kohn2008}.
Rather than mapping a single point to the inner boundary of the cloak Kohn et al. proposed mapping a ball of small, but finite, radius to the inner boundary~\cite{Kohn2008}.
A small regularisation parameter which characterises the initial radius of the ball is introduced, which results in a non-singular mapping on the cloak and its boundary.
The regularisation procedure was used to create illustrative \emph{near cloaks} in~\cite{Norris2008}.

In 2006, Milton et al.~\cite{Milton2006} examined how the equations of motion for a general elastic medium transform under an arbitrary curvilinear transformation.
It was shown that a priori requiring a symmetric stress tensor enforces a particular choice of the gauge (i.e. the manner in which the displacements transforms).
It was found that, in general, the equations of motion are not invariant under transformation but are mapped to a more general system with non-scalar density.
Milton et al. demonstrated that a special case of the so-called Willis equations~\cite{Milton2007} remain invariant under general curvilinear transformations.
In~\cite{Milton2006} identities linking the material properties of the transformed material to the transformation, for both classical elasticity and the more general Willis materials are derived.
In 2011, Norris et al.~\cite{Norris2011} further generalised the work of Milton et al., deriving a more general system of transformed equations without imposing the constraint of symmetric stress.
The material properties of the transformed system were derived explicitly and shown to depend on both the transformation and choice of gauge.
Together~\cite{Milton2006} and~\cite{Norris2011} provide a comprehensive framework in which to investigate cloaking in elastodynamics.

A design for a cloak to control flexural waves in thin plates was proposed by Farhat et al.~\cite{Farhat2009}.
The cloak is constructed of several concentric layers of piecewise constant isotropic elastic material.
Farhat et al. also presented a simplified model suitable for practical implementation with ten layers using six different materials.
Following~\cite{Farhat2009}, an experimental group led by Wegener fabricated a cloak based on the work of Farhat et al. using twenty concentric rings and sixteen different elastic metamaterials~\cite{Stenger2012}.
Physical measurements were compared with numerical simulations and found to be in good agreement.
Control of in-plane waves governed by the Navier equations was examined by Brun et al.~\cite{Brun2009}.
In~\cite{Brun2009}, the authors modelled a circular cloak using the classical radial transformation by deforming a disc to an annulus.
The efficiency of the cloak was illustrated using finite element simulations and the numerical solution of the cloaking problem was compared with the Green's function for a homogeneous elastic space.

\begin{figure}[htb]
\centering
\begin{subfigure}[c]{0.4\textwidth}
\includegraphics[width=\linewidth]{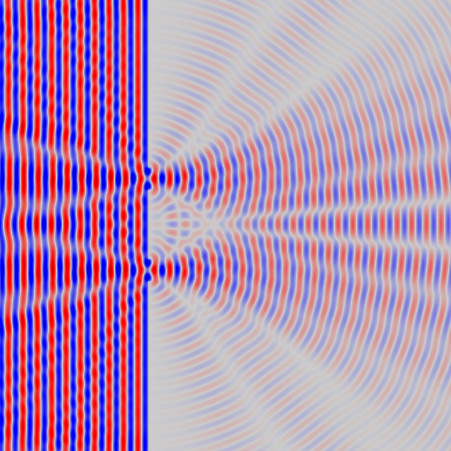}
\caption{\label{fig:Youngs-Intact}}
\end{subfigure}
\qquad
\begin{subfigure}[c]{0.4\textwidth}
\includegraphics[width=\linewidth]{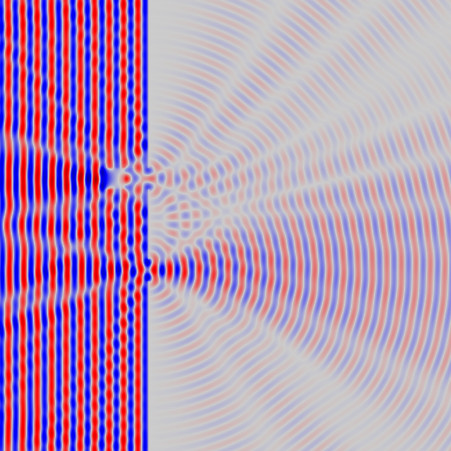}
\caption{\label{fig:Youngs-Uncloaked}}
\end{subfigure}
\qquad

\begin{subfigure}[c]{0.4\textwidth}
\includegraphics[width=\linewidth]{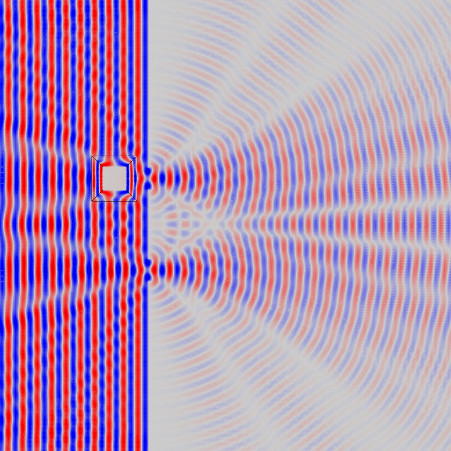}
\caption{\label{fig:Youngs-Cloaked}}
\end{subfigure}
\begin{subfigure}[c]{0.45\textwidth}
\vspace{0.06\textwidth}
\includegraphics[width=\linewidth]{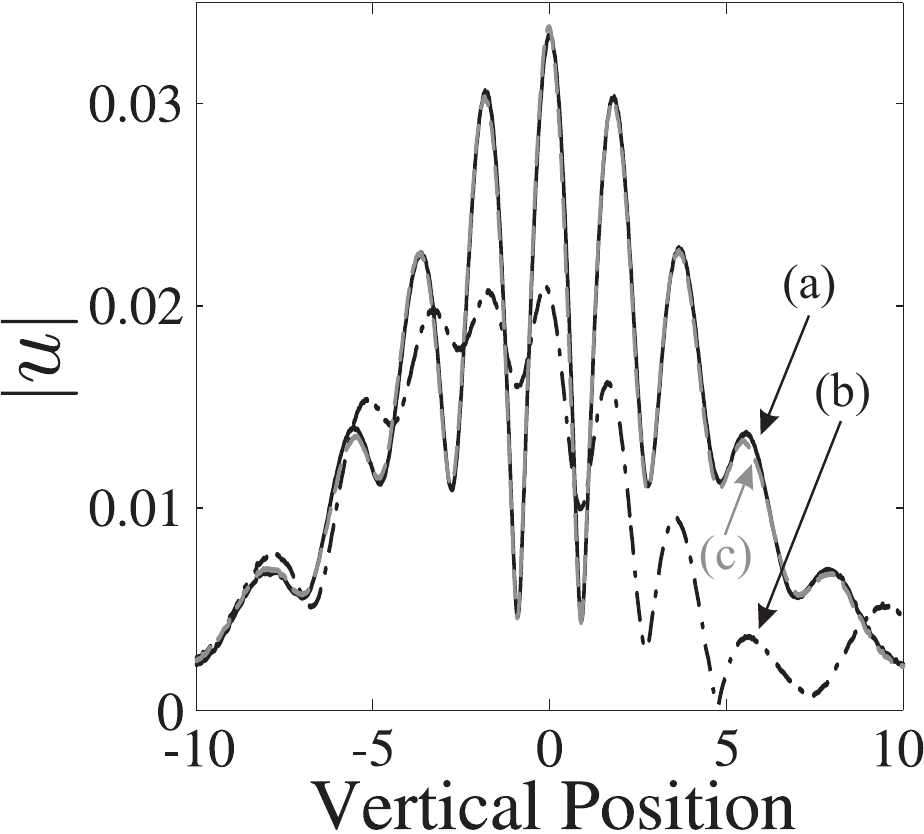}
\hspace{0.05\textwidth}
\caption{\label{fig:Youngs-Comp}}
\end{subfigure}
\caption{\label{fig:youngs-slits}
(a)-(c) The field $u(\vec{x})$ for the Young's double slit experiment with no inclusion, an uncloaked inclusion, and a cloaked inclusion respectively.
(d) A plot of $|u(\vec{x})|$ over the observation screen illustrating the interference fringes for cases (a)-(c).
(An animated version of this figure may be found in the supplementary material~\cite{ESM}.)}
\end{figure}

An influential paper \cite{Rahm2008} presents a transformation optics algorithm for a problem of electromagnetism involving a cloak of a square shape.
The transformation is performed in Cartesian coordinates  and results in a piecewise smooth cloak on the interior points, with matching regions in the neighbourhood of corners and a singularity at the origin transformed into the inner boundary of the cloak.
The model of such a continuum cloak received substantial attention and subsequent use by the modelling community (see, for example, \cite{Farhat2008,Savic2012,Gallina2010,Li2012,Paul2012,Jiang2009}).
In the majority of these papers, the emphasis is on the geometrical aspect of the possible shapes of the cloak, with examples ranging from polygonal and elliptical cloaks to heart-shaped cloaks. Although it is indeed interesting to see a wide range of transformations and geometries, it also remains important to understand the transformed problem in the context of the physical model, address the analysis of the transformed boundary or transmission conditions and furthermore derive the properties of the solutions. The paper \cite{Rahm2008}, which stimulated a good level of discussion on the topic, also admits a deficiency regarding the analysis of the solution near the boundary of the cloak. 
Apparently, no indication is given about the sensitivity of the result to the type of boundary conditions (Dirichlet or Neumann) on the inner boundary of the cloak. 
The authors' evaluation of the effectiveness of the cloaking is based on a visual observation linked to a numerical simulation at a single frequency. 
On page 91, the text of \cite{Rahm2008} says ``Although the effective $\mu_x$ is inaccurate in the vicinity of 
the boundary of the inner square, it can be shown, that the performance of the implemented device is not affected by this fact, which is out of the scope of this paper.''
Indeed, if the authors had attempted to change the frequency range they would have seen significant differences. Hence, the cloak advocated in \cite{Rahm2008} is an approximate cloak, where the boundary effects become important and visible as the frequency of the incident waves increases. 

The ideas of metric invariance in Maxwell's equations and cloaking have been taken on board as a technical tool and on many occasions,  the researchers omit  to look at the physical model corresponding to the transformed equations.
For example, on page 99 in \cite{Li2012} the text reads ``The square cloak has the same geometry as the cylindrical case, except that we replace the cylindrical shell by a rectangular shell with the same size''.
This comparison of unlike geometries omits important effects, such as field concentrations near sharp corners, which make cloaking more difficult.
Motivated by~\cite{Rahm2008}, Farhat et al.~\cite{Farhat2008} attempted to construct an approximate square cloak for out-of-plane shear waves.
Farhat et al., using the method of multiple scales, introduced a microstructure composed of a regular array of perforations and derived a homogenised continuum which would approximate an ideal cloak.
However, as Farhat et al. admit on page 15 of~\cite{Farhat2008} ``it is fair to say that our structured cloak is not as efficient as we would have expected''.

Polygonal cloaks have also been the subject of experimental investigation.
For example, Chen et al. report~\cite{Chen2012} the results of an experimental broadband hexagonal cloak based on a piecewise linear homogeneous transformation.
Although the cloak does not render the cloaked object invisible, it does reduce its apparent size.
The cloak is demonstrated to work for visible light.
However, Chen et al. emphasise that the cloak only functions for light incident from six directions defined by the faces of the hexagon.
More recently, Landy et al. produced an experimental uni-directional metamaterial cloak for microwaves.
The reported cloak~\cite{Landy2012} is based on a bilinear transformation which maps a line segment to a two dimensional region of space.
Cloaks based on such transformations are referred to as \emph{carpet cloaks} in the literature (see~\cite{Li2008} among others).
The advantage of such cloaks is that the requisite material properties are homogeneous and finite.
However, Landy et al. admit that such cloaks are only effective over a narrow range of observation angles.
The cloak is nonetheless impressive given that the practical implementation does not rely on the eikonal approximations as is the case with other implementations~\cite{Chen2011,Zhang2011,Popa2011}.

It appears that the work reported in~\cite{Rahm2008} has generated a scope for further discussion and indeed further improvement of the model involving ``glued transforms" that lead to approximate rectangular cloaks.
Such cloaks are by no means exact and are frequency sensitive.
A regularisation procedure, as illustrated by Kohn et al. for the spherical cloak~\cite{Kohn2008}, can be applied to make the transformation, and hence the material properties, non-singular on the inner boundary of the cloak.
The regularisation procedure not only simplifies the analysis, but also makes it physically meaningful.
Furthermore, a lattice approximation is straightforward for a regularised square-shaped cloak.
This appears to be efficient and serves a relatively wide frequency range.

It should be emphasised that the current paper improves significantly the influential work by Rahm et al.~\cite{Rahm2008}.
In particular, the square \emph{push out} transformation is regularised and it is demonstrated that this construction of a \emph{near cloak} via regularisation does not adversely affect the effectiveness of the cloaking.
Detailed ray analysis is carried out and the ray equations are derived.
An objective numerical measure of the quality of the cloaking is introduced and used to quantify the effects of frequency, interface conditions, and orientation of the inclusion.
Illustrative simulations are produced for cylindrical sources and the analytical ray diagrams presented.
For physical fabrication, a discrete lattice cloak is used to approximate the regularised continuum transform and it is shown that the lattice structure acts as an effective cloak, particularly at low frequencies.

In the present paper, the emphasis is on solutions of the Helmholtz equation.
The Helmholtz equation arises in a wide variety of fields including electromagnetism, elasticity, and acoustics.
Therefore, solutions to the cloaking problem for the Helmholtz equation have a wide range of potential applications.
However, for definiteness and ease of exposition the language of elasticity will be used throughout this paper.

The current paper is structured as follows.
A description of the regularised cloak in the continuum model of out-of-plane shear elastic waves follows the introduction.
This also includes the  discussion of the natural interface conditions on the boundaries of the cloak.
An explicit ray tracing algorithm is developed, and the phenomenon of negative refraction on the interface boundaries is explained.
Numerical scattering measures are included, with detailed simulations.
The analysis also incorporates Neumann and Dirichlet boundary conditions on the inner contour of the cloak.
As a demonstration of the effectiveness of the regularised cloak, a Young's double slit experiment is presented.
To give the reader an idea of the results which follow, we show part of a simulation of the interference pattern due to a plane wave interacting with two apertures in figure~\ref{fig:youngs-slits}.
The same figure presents the fringe structure for the case of an aperture obstructed  by an uncloaked and then cloaked inclusion.
This numerical experiment convincingly demonstrates the high quality of the cloaking effect produced by this regularised cloak.
Further discussion of the Young's double slit experiment is deferred to section~\ref{sec:Youngs}.
It is also shown that one of the undeniable advantages of such an approximate cloak is the straightforward connection with the discrete lattice structures.
These connections are analysed in detail, and accompanied by a range of physical simulations.
Concluding discussion and a final outline are included in the last section of the paper.

\section{The regularised continuum cloak}

The classical approach to cloaking via transformation geometry involves deforming a region such that a point is mapped to a finite region corresponding to the inner boundary of the cloak.
Indeed, the square \emph{push out} transformation proposed by Rahm et al.~\cite{Rahm2008} maps a point to a square.
The mapping is non-singular everywhere except at the inner boundary of the cloak.
In the present paper, a regularised version of the square \emph{push out} transformation is used.
In particular, the trapezoids $\chi^{(i)}$ are mapped to the trapezoids $\Omega_-^{(i)}$ as illustrated in figure~\ref{fig:deformation} with continuity, but not smoothness, imposed on the interfaces between the four trapezoids.
The mapping is non-singular on the closure of the cloak, and hence, all corresponding material properties are finite.
It will be shown that this regularised transformation yields an effective broadband cloak, with finite material properties which may easily be approximated by a regular lattice.

\subsection{The transformation}
\label{subsec:transformation}

\begin{figure}[htb]
\centering
\begin{tikzpicture}
\begin{scope}[shift={(-4,-2)}]
\draw[] decorate [decoration={random steps}] {(-3.25,-3.25) rectangle (3.25,3.25)};
\draw[fill=lightgray] (-2,-2) rectangle (2,2);
\draw[fill=white] (-0.2,-0.2) rectangle (0.2,0.2);
\draw (0.2,0.2) -- (2,2);
\draw (-0.2,0.2) -- (-2,2);
\draw (-0.2,-0.2) -- (-2,-2);
\draw (0.2,-0.2) -- (2,-2);
\draw[->, bend left,line width = 1] (-2.2,-2.2) to (0,0);
\node[below] at (-2.2,-2.2) {$\displaystyle \chi_0$};
\node at (1.1,0) {$\displaystyle \chi^{(1)}$};
\node at (0,1.1) {$\displaystyle \chi^{(2)}$};
\node at (-1.1,0) {$\displaystyle \chi^{(3)}$};
\node at (0,-1.1) {$\displaystyle \chi^{(4)}$};
\node[below] at (0,3) {$\displaystyle \Omega_+$};
\node[right] at (2,-1.5) {$\displaystyle \Gamma^{(1)}$};
\node[above] at (1.5,2) {$\displaystyle \Gamma^{(2)}$};
\node[left] at (-2,1.5) {$\displaystyle \Gamma^{(3)}$};
\node[below] at (-1,-2) {$\displaystyle \Gamma^{(4)}$};
\end{scope}
\begin{scope}[shift={(4,2)}]
\draw[] decorate [decoration={random steps}] {(-3.25,-3.25) rectangle (3.25,3.25)};
\draw[fill=lightgray] (-2,-2) rectangle (2,2);
\draw[fill=white] (-1.5,-1.5) rectangle (1.5,1.5);
\draw (1.5,1.5) -- (2,2);
\draw (-1.5,1.5) -- (-2,2);
\draw (-1.5,-1.5) -- (-2,-2);
\draw (1.5,-1.5) -- (2,-2);
\node at (0,0) {$\displaystyle \Omega_0$};
\node[right,fill=white] at (3,0) {$\displaystyle \Omega_-^{(1)}$};
\node[above,fill=white] at (0,3) {$\displaystyle \Omega_-^{(2)}$};
\node[left,fill=white] at (-3,0) {$\displaystyle \Omega_-^{(3)}$};
\node[below,fill=white] at (0,-3) {$\displaystyle \Omega_-^{(4)}$};
\draw[->, bend left,line width = 1] (3,0) to (1.75,0);
\draw[->, bend left,line width = 1] (0,3) to (0,1.75);
\draw[->, bend left,line width = 1] (-3,0) to (-1.75,0);
\draw[->, bend left,line width = 1] (0,-3) to (0,-1.75);
\node[right] at (2,-1.5) {$\displaystyle \Gamma^{(1)}$};
\node[above] at (1.5,2) {$\displaystyle \Gamma^{(2)}$};
\node[left] at (-2,1.5) {$\displaystyle \Gamma^{(3)}$};
\node[below] at (-1.5,-2) {$\displaystyle \Gamma^{(4)}$};
\node[] at (1.2,0) {$\displaystyle \gamma^{(1)}$};
\node[] at (0,1.2) {$\displaystyle \gamma^{(2)}$};
\node[] at (-1.2,0) {$\displaystyle \gamma^{(3)}$};
\node[] at (0,-1.2) {$\displaystyle \gamma^{(4)}$};
\node[below] at (-2.5,3) {$\displaystyle \Omega_+$};
\end{scope}
\draw[->, bend right,line width = 2] (0,-4) to (3,-2);
\node at (1.4,-3) {$\displaystyle \vec{\mathcal{F}}$};
\end{tikzpicture}
\caption{\label{fig:deformation}
The map $\vec{\mathcal{F}}$ maps the undeformed region $\chi$ to the deformed configuration $\Omega_-$.
The boundary between $\Omega_+$ and $\Omega_-^{(i)}$ is denoted $\Gamma^{(i)}$, while the interface between $\Omega_0$ and $\Omega_-^{(i)}$ is denoted $\gamma^{(i)}$.
The corresponding boundaries in the undeformed configuration are denoted by $\Gamma^{(i)}$ and $\sigma^{(i)}$ respectively.
}
\end{figure}
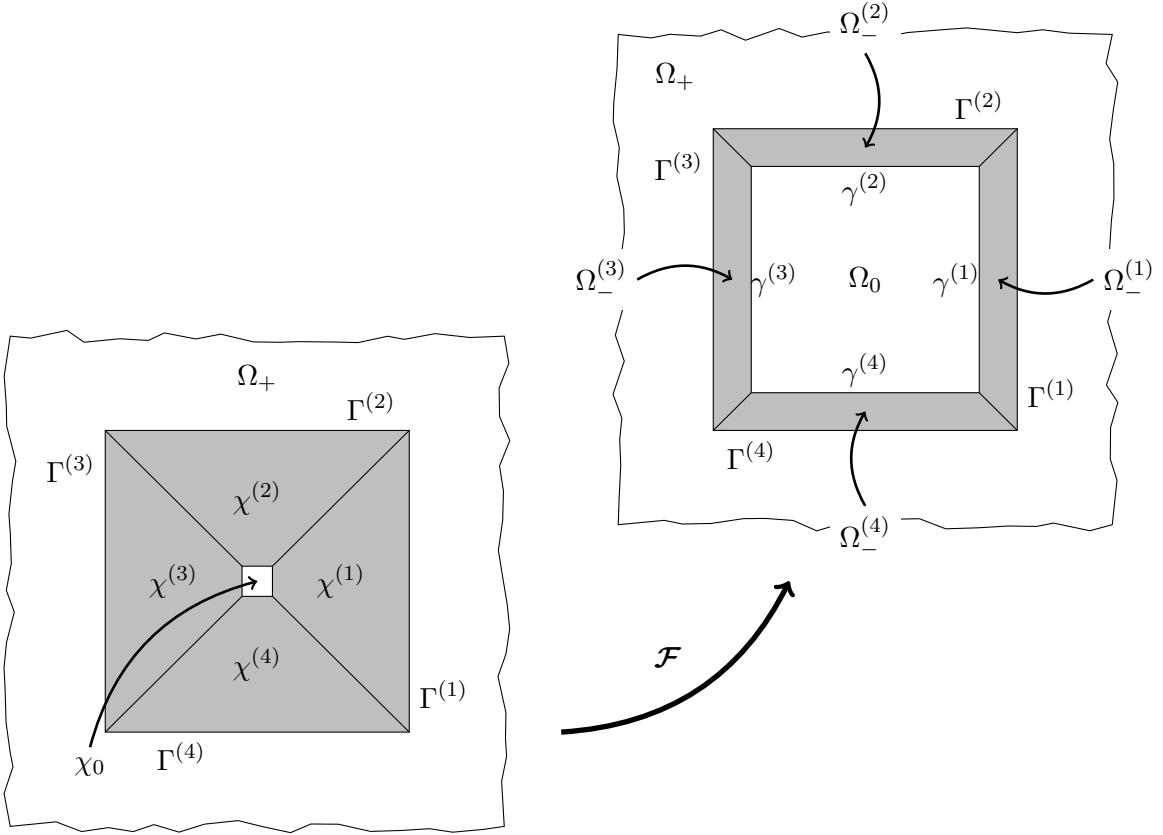

Consider a small square $\chi_{0} = \{\vec{X}:|X_{1}| < \epsilon, |X_{2}| < \epsilon\}\subset\mathbb{R}^{2}$, which via the map $\vec{\mathcal{F}}$ is mapped to the square $\Omega_{0} =  \{\vec{x}:|x_{1}| < a,  |x_{2}| < a\}\subset\mathbb{R}^{2}$.
Physically $w$ is the thickness of the cloak, $a$ is the semi-width of the inclusion $\Omega_0$, and $\epsilon$ is the initial semi-width of the square $\chi_0$ where $0<\epsilon/a\ll 1$.
In this case, it is convenient to decompose the cloak into four sub-domains $\chi=\chi^{(1)}\cup\ldots\cup\chi^{(4)}$, as illustrated in figure~\ref{fig:deformation}.
Formally, the map $\vec{\mathcal{F}}$ defines a pointwise map from $\vec{X}\in\chi=\chi^{(1)}\cup\ldots\cup\chi^{(4)}$ to $\vec{x}=\vec{\mathcal{F}}(\vec{X})\in\Omega_-=\Omega_-^{(1)}\cup\ldots\cup\Omega_-^{(4)}$.
The mapping is continuous and non-linear on $\chi$, and defined in a piecewise fashion such that $\vec{\mathcal{F}} = \vec{\mathcal{F}}^{(i)}(\vec{X})$ for $\vec{X}\in\chi^{(i)}$, where
\[
\vec{\mathcal{F}}^{(1)}(\vec{X}) = \begin{bmatrix}
\alpha_1 X_1 + \alpha_2 \\
\alpha_1 X_2 + \alpha_2X_2/X_1
\end{bmatrix},\qquad
\vec{\mathcal{F}}^{(2)}(\vec{X}) = \begin{bmatrix}
\alpha_1 X_1 + \alpha_2X_1/X_2 \\
\alpha_1 X_2 + \alpha_2
\end{bmatrix}
\]
\[
\vec{\mathcal{F}}^{(3)}(\vec{X}) = \begin{bmatrix}
\alpha_1 X_1 - \alpha_2 \\
\alpha_1 X_2 - \alpha_2X_2/X_1
\end{bmatrix},\qquad
\vec{\mathcal{F}}^{(4)}(\vec{X}) = \begin{bmatrix}
\alpha_1 X_1 - \alpha_2 X_1/X_2 \\
\alpha_1 X_2 - \alpha_2
\end{bmatrix},
\]
with $\alpha_1 = w/(a+w-\epsilon)$ and $\alpha_2 = (a+w)(a-\epsilon)/(a+w-\epsilon)$.
The exterior of the cloak remains unchanged by the map, that is, $\vec{X} = \vec{\mathcal{F}}(\vec{X})$ for $\vec{X}\in\bar{\Omega}_+$, where the bar denotes the closure of the domain.
The Jacobian matrices and determinants are then
\[
\vec{J}^{(1)} = \begin{pmatrix}
\alpha_1 & 0 \\
\\
\dfrac{x_2\alpha_1\alpha_2}{x_1(\alpha_2-x_1)} & \dfrac{x_1 \alpha_1}{x_1-\alpha_2}
\end{pmatrix},\qquad
\vec{J}^{(2)} = \begin{pmatrix}
\dfrac{x_2\alpha_1}{x_2-\alpha_2} & \dfrac{x_1\alpha_1\alpha_2}{x_2(\alpha_2-x_2)} \\
\\
0 & \alpha_1
\end{pmatrix},
\]
\[
\vec{J}^{(3)} = \begin{pmatrix}
\alpha_1 & 0 \\
\\
\dfrac{x_2\alpha_1\alpha_2}{x_1(\alpha_2+x_1)} & \dfrac{x_1 \alpha_1}{x_1+\alpha_2}
\end{pmatrix},\qquad
\vec{J}^{(4)} = \begin{pmatrix}
\dfrac{x_2\alpha_1}{x_2+\alpha_2} & \dfrac{x_1\alpha_1\alpha_2}{x_2(\alpha_2+x_2)} \\
\\
0 & \alpha_1
\end{pmatrix},
\]
\[
J^{(1)} = \frac{x_1\alpha_1^2}{x_1-\alpha_2},\qquad
J^{(2)} = \frac{x_2\alpha_1^2}{x_2-\alpha_2},\qquad
J^{(3)} = \frac{x_1\alpha_1^2}{x_1+\alpha_2},\qquad
J^{(4)} = \frac{x_2\alpha_1^2}{x_2+\alpha_2}.
\]
It is emphasised that $J^{(i)}(x_{i})=\det\vec{J}^{(i)}\neq 0$ for $\vec{x}\in\bar{\Omega}_{-}^{(i)}$ and $\epsilon\neq0$, that is, the map is continuous on both the interior and boundary of the cloak.
The metric of the deformed space $\Omega_{-}^{(i)}$ is $\vec{g}^{(i)} = (\vec{J}^{(i)}{\vec{J}^{(i)}}^\text{T})^{-1}$.

The present paper will be devoted to the propagation of time harmonic out-of-plane shear waves of radian frequency $\omega$ and displacement amplitude $u(\vec{x})$.
Lemma 2.1 in~\cite{Norris2008} allows the Helmholtz equation for an isotropic homogeneous medium $\mu\nabla_{\vec{X}}\cdot(\nabla_{\vec{X}}) u(\vec{X}) + \varrho\omega^2u(\vec{X}) = 0$ for $\vec{X}\in\chi$ to be written in deformed co-ordinates as
\begin{equation}
[\nabla\cdot(\vec{C}^{(i)}(\vec{x})\nabla) + \rho^{(i)}(\vec{x})\omega^2] u(\vec{x}) = 0,
\qquad \vec{x}\in\Omega_-^{(i)},
\end{equation}
where $\mu$ is the constant ambient stiffness, $\varrho$ is the constant ambient density, $\vec{C}^{(i)}(\vec{x}) = [\mu/J^{(i)}(\vec{x})]\vec{J}^{(i)}(\vec{x})[\vec{J}^{(i)}(\vec{x})]^\mathrm{T}$ is the transformed stiffness matrix and $\rho^{(i)}(\vec{x}) = \varrho/J^{(i)}(\vec{x})$ is the transformed density.
The differential operator $\nabla_{\vec{X}}$ is written in the undeformed space and should be distinguished from $\nabla$ which is written in the deformed coordinates.

Since the mapping is continuous on $\bar{\Omega}_-$, the material properties of the cloak are non-singular.
The transformed stiffness tensor is symmetric and positive definite.
The density is a scalar and bounded on $\bar{\Omega}_-$.
Physically, the transformed material properties correspond to a heterogeneous anisotropic medium.

\subsection{Interface conditions}
\label{sec:interface}
Without loss of generality, it is convenient to restrict the following analysis to a single side of the cloak.
With reference to figure~\ref{fig:deformation}, consider the sub-domain $\Omega_-^{(1)}\subset\mathbb{R}^2$ in the absence of the inclusion and remaining three sides of the cloak.
In the absence of sources the out-of-plane shear deformation amplitude of an outgoing time-harmonic wave of angular frequency $\omega$ satisfies the following equation
\begin{equation}
\mathcal{L}u(\vec{x}) = 0,
\label{eq:cont-helm}
\end{equation}
together with the Sommerfield radiation condition at infinity.
Here $\mathcal{L} = \nabla\cdot(\vec{A}(\vec{x})\nabla) + \rho(\vec{x})\omega^2$ is the Helmholtz operator, $\vec{A}(\vec{x})$ and $\rho(\vec{x})$ are defined as
\begin{equation}
\vec{A}(\vec{x}) = \begin{cases}
\vec{C}^{(1)}(\vec{x}) & \text{for }\vec{x}\in\Omega_-^{(1)} \\
\mu\mathbb{I} & \text{for }\vec{x}\in\Omega_+
\end{cases}, \qquad
\rho(\vec{x}) = \begin{cases}
\rho^{(1)}(\vec{x}) & \text{for }\vec{x}\in\Omega_-^{(1)} \\
\varrho & \text{for }\vec{x}\in\Omega_+
\end{cases}.
\label{eq:A-and-rho-def}
\end{equation}

Let $v(\vec{x})$ be a continuous piecewise smooth solution of the Helmholtz equation in $\mathbb{R}^2$ satisfying the Sommerfield radiation condition at infinity.
Integrating the difference $u(\vec{x})\mathcal{L}v(\vec{x}) - v(\vec{x})\mathcal{L}u(\vec{x})$ over a disc $\mathcal{D}_r$ of radius $r$ containing $\Omega^{(1)}_{-}$ yields
\[\begin{split}
0 & = \int\limits_{\mathcal{D}_r} \left(
u\nabla\cdot\vec{A}\nabla v - v\nabla\cdot\vec{A}\nabla u
\right) \dd\vec{x},\\
& = \int\limits_{\partial\Omega_-^{(i)}} \left(
u^-\vec{n}\cdot\vec{A}\nabla v^- - v^-\vec{n}\cdot\vec{A}\nabla u^-\right) \dd\vec{x} -
\int\limits_{\partial\Omega_-^{(i)}} \left(
u^+\vec{n}\cdot\vec{A}\nabla v^+ - v^+\vec{n}\cdot\vec{A}\nabla u^+\right) \dd\vec{x}\\
& \hspace{5em}
+ \mu\int\limits_{\partial\mathcal{D}_r} \left(u\vec{n}\cdot\nabla v +v\vec{n}\cdot\nabla u\right)\, \dd\vec{x},
\end{split}
\]
where the fact that $\nabla u\cdot\vec{A}\nabla v = \nabla v\cdot\vec{A}\nabla u$ (since $\vec{A}$ is symmetric) has already been used.
Since $u(\vec{x})$ and $v(\vec{x})$ represent outgoing solutions, the final integral vanishes as $r\to\infty$.
Thus, the essential interface condition is continuity of the field
\begin{equation}
\left[u\right] = 0 \qquad\text{on} \qquad \partial\Omega_{-}^{(1)},
\label{eq:cont-field}
\end{equation}
and the natural interface condition is continuity of tractions
\begin{equation}
\vec{n}\cdot\vec{C}^{(1)}\nabla u^{-} = \mu\vec{n}\cdot\nabla u^{+} \qquad\text{on} \qquad \partial\Omega_{-}^{(1)}.
\label{eq:cont-trac}
\end{equation}
These interface conditions are equivalent to those derived in~\cite{Norris2008} for acoustic pressure by application of Nanson's formula and imposing particle continuity.

\subsection{The cloaking problem}

Consider the propagation of time harmonic out-of-plane deformations, generated by a point source, in a homogeneous infinite elastic solid in which is embedded an inclusion surrounded by a cloak.
The displacement amplitude field then satisfies
\begin{equation}
[\nabla\cdot(\vec{A}(\vec{x})\nabla) + \rho(\vec{x})\omega^{2}]u(\vec{x}) = -\delta(\vec{x}-\vec{x}_{0}), \qquad \vec{x}\in\mathbb{R}^{2}\setminus\bar{\Omega}_{0},
\qquad \vec{x}_0 \in\Omega_+
\end{equation}
\begin{equation}
[\mu_0\nabla\cdot(\nabla) + \varrho_{0}\omega^{2}]u(\vec{x}) = 0, \qquad \vec{x}\in\Omega_{0},
\end{equation}
with continuity of $u(\vec{x})$ and tractions on all internal boundaries according to~\eqref{eq:cont-field} and~\eqref{eq:cont-trac}.
Additionally, the Sommerfeld radiation condition is imposed at infinity.
The stiffness tensor $\vec{A}(\vec{x})$ and density $\rho(\vec{x})$ are 
\begin{equation}
\vec{A}(\vec{x}) = \begin{cases}
\vec{C}^{(i)}(\vec{x}) & \text{for }\vec{x}\in\Omega_-^{(i)} \\
\mu\mathbb{I} & \text{for }\vec{x}\in\Omega_+
\end{cases}, \qquad
\rho(\vec{x}) = \begin{cases}
\rho^{(i)}(\vec{x}) & \text{for }\vec{x}\in\Omega_-^{(i)} \\
\varrho & \text{for }\vec{x}\in\Omega_+
\end{cases},
\end{equation}
and $\mu_{0}$ and $\rho_{0}$ are the stiffness and density of the inclusion respectively.

\subsection{The ray equations}

Whilst, in principle, the exact wave behaviour of the displacement field can be obtained from the solution of the cloaking problem, it is illuminating to consider the leading order behaviour of rays through the cloak.
Consider a WKB expansion of the displacement amplitude field in terms of angular frequency $\omega
$, and the amplitude and phase functions $U_n(\vec{x})$ and $\varphi(\vec{x})$ respectively
\begin{equation}
u(\vec{x}) \sim e^{i\omega\varphi(\vec{x})}\sum_{n=0}^{\infty} \frac{i^{n} U_{n}(\vec{x})}{\omega^{n}},\qquad\text{as}\;\omega\to\infty,
\end{equation}
whence the leading order equation for the phase on the interior of the cloak is
\begin{equation}
H(\vec{x},\vec{s}) = 0,
\label{eq:hamiltonian}
\end{equation}
where $H(\vec{x},\vec{s}) = \mu\varrho^{-1}\vec{s}\cdot\vec{g}^{-1}\vec{s} - 1$, $\vec{s} = \nabla\varphi$ is the slowness vector, $\mu$ and $\varrho$ are the stiffness and density of the ambient medium respectively, and $\vec{g}$ is the metric of the transformation.
In terms of wave propagation, the conserved quantity $H(\vec{x},\vec{s})$ represents the first order slowness contours~\cite{Musgrave}.
The characteristics of the quantity $H(\vec{x},\vec{s})$ then satisfy the following system
\begin{equation}
\frac{\dd H}{dt} = 0,\qquad
\frac{\dd \vec{x}}{\dd t} = \frac{\partial H}{\partial\vec{s}},\qquad
\frac{\dd \vec{s}}{\dd t} = -\frac{\partial H}{\partial\vec{x}},
\label{eq:characteristic-system}
\end{equation}
where $t$ is the ray parameter (equivalently, the time-like parameter).
At this point, it is convenient to introduce index summation notation where summation, from 1 to 2, over repeated indices is implied.
The system~\eqref{eq:characteristic-system} may then be expressed as
\begin{equation}
\frac{\dd s_{i}}{\dd t} = - 2\varrho^{{-1}}\mu s_{m}s_{n}J_{nl}\frac{\partial J_{ml}}{\partial x_{i}}, \qquad
\frac{\dd x_{i}}{\dd t} = 2\varrho^{{-1}}\mu J_{il}J_{jl}s_{j},
\label{eq:cloak-char}
\end{equation}
where $J_{ij} = (\vec{J})_{ij}$ are the components of the Jacobian matrix and should be distinguished from the $J$, the Jacobian determinant.
The superscript labels have been omitted for brevity, but $J_{ij}$ and $J$ should be understood as $J^{(k)}_{ij}$ and $J^{(k)}$ for $k=1,\ldots4$ corresponding to the four sides of the cloak.

From equation~\eqref{eq:hamiltonian}, an alternative representation, in terms of wave normals $\vec{n}$ and the phase velocity $v$, is
\begin{equation}
\mu\varrho^{-1}\vec{n}\cdot\vec{g}^{-1}\vec{n} - v^{2} = 0.
\label{eq:hamiltonian2}
\end{equation}
The representation~\eqref{eq:hamiltonian2} is obtained by assuming a plane wave solution to the Helmholtz equation (see, for example~\cite{Musgrave}).
Alternatively, seeking a solution of the full wave equation in the form of the leading term in a WKB expansion
\[
u(\vec{x},t) \sim e^{i\omega\varphi(\vec{x},t)}\sum_{n=0}^{\infty} \frac{i^{n} U_{n}(\vec{x},t)}{\omega^{n}},\qquad\text{as}\;\omega\to\infty,
\]
yields the same result with $\partial \varphi/\partial t = v$.
From~\eqref{eq:hamiltonian} and~\eqref{eq:hamiltonian2} the slowness vector can be expressed in terms of the original material properties (through $\varrho$ and $\mu$) and the map (through $\vec{J}$) as
\begin{equation}
\vec{s} = \frac{\vec{n}}{v} = \frac{\vec{n}}{|\vec{J}^\mathrm{T} \vec{n}|}\sqrt{\frac{\varrho}{\mu}},
\label{eq:s-S}
\end{equation}
Further, in the undeformed configuration, the equivalent conserved quantities are $\mu\varrho^{-1}\vec{S}\cdot\vec{S} - 1 = 0$ and $\mu\varrho^{-1} = V^{2}$.
Together with~\eqref{eq:hamiltonian} and~\eqref{eq:hamiltonian2}, these two equations imply that
\begin{equation}
\vec{s} = \vec{J}^{-\mathrm{T}}\vec{S} = \frac{\vec{J}^{-\mathrm{T}}\vec{N}}{V} = \vec{J}^{-\mathrm{T}}\vec{N}\sqrt{\frac{\varrho}{\mu}}.
\label{eq:s-N}
\end{equation}

Now, consider a ray (line) in the ambient medium, in direction $\vec{N}$ passing through $\vec{X}_{0}$ and parameterised by $t$.
The corresponding curve in the cloak is $\vec{x}(t) = \vec{\mathcal{F}}(\vec{X}_{0}+t\vec{N})$, whence
\[
\frac{\dd x_i}{\dd t} = J_{ij}N_{j}, 
\]
which using~\eqref{eq:s-N} may be rewritten thus
\begin{equation}
\frac{\dd x_i}{\dd t} = J_{il}J_{jl}s_j\sqrt{\frac{\mu}{\varrho}}.
\label{eq:dx-ray}
\end{equation}
Taking the derivative of~\eqref{eq:s-N} for constant $\vec{N}$ yields
\[
\frac{\dd s_i}{\dd t} =
 s_k s_n J_{kj}J_{lm}J_{nm}\frac{\partial {J^{-1}}_{ji} }{\partial x_l}\sqrt{\frac{\mu}{\varrho}}.
\]
Using the compatibility condition that the deformation gradient should be irrotational under finite deformation $\epsilon_{jk}\partial {J^{-1}}_{ik}/\partial x_j = 0_{i}$, the partial derivative above may be written as $\partial {J^{-1}}_{jl}/\partial x_i$, whence
\begin{equation}
\frac{\dd s_i}{\dd t} = - s_m s_n J_{n\ell}\frac{\partial J_{ml}}{\partial x_i}\sqrt{\frac{\mu}{\varrho}},
\label{eq:ds-ray}
\end{equation}
where $\epsilon_{jk}$ is the permutation tensor and the equality $J_{lm}\partial J^{-1}_{jl}/\partial x_{i} = - J^{-1}_{jl}\partial J_{lm}/\partial x_{i}$ has already been used.
Consider the characteristic equations for the waves in the cloak~\eqref{eq:cloak-char}, together with the equations of the transformed rays~\eqref{eq:dx-ray} and~\eqref{eq:ds-ray}.
The system~\eqref{eq:dx-ray} and~\eqref{eq:ds-ray} are the equations of characteristics in the cloak, up to an arbitrary scaling constant of $2\sqrt{\mu/\varrho}$.
Thus, to leading order, rays (or straight lines) in the ambient medium map directly to rays in the cloak.

\begin{figure}[htb]
\centering
\begin{subfigure}[c]{0.4\textwidth}
\includegraphics[width=\textwidth]{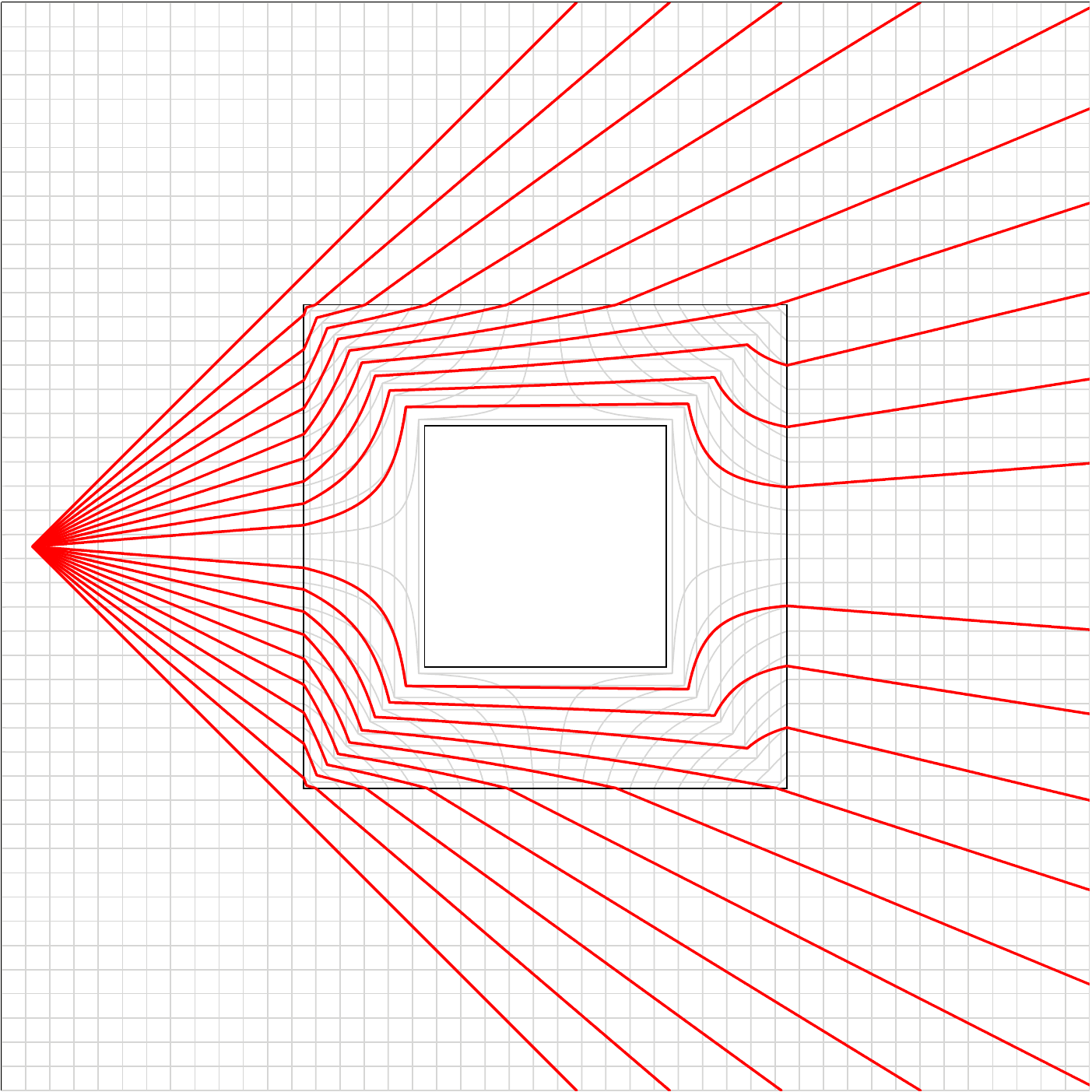}
\caption{\label{fig:Rays-Side}}
\end{subfigure}
\qquad
\begin{subfigure}[c]{0.4\textwidth}
\includegraphics[width=\textwidth]{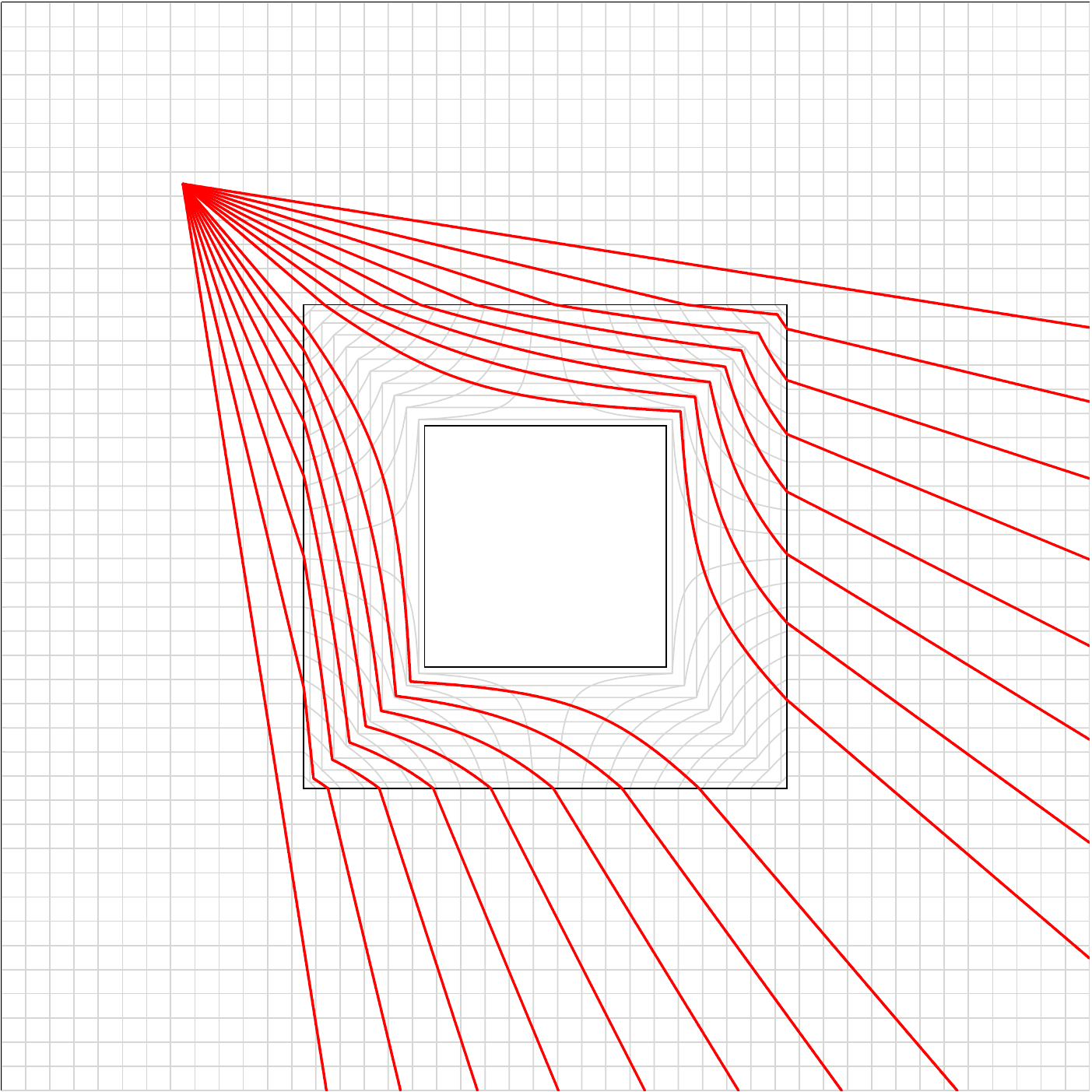}
\caption{\label{fig:Rays-Corner}}
\end{subfigure}
\caption{\label{fig:Rays}
Plots of the ray paths through the cloak for a cylindrical source.
The grey lines indicate the deformation of the space inside the cloak.
(Animated versions of these figures may be found in the supplementary material~\cite{ESM}.)}
\end{figure}

Figure~\ref{fig:Rays} shows rays emanating from a point source, passing through the cloak and emerging from the cloak on their original trajectory.
In this sense, the object is ``invisible'' to an observer outside the cloaking region.
The figure clearly illustrates how wave propagation in the cloak is related to the map.
Animated versions of figure~\ref{fig:Rays} can be found in the supplementary material~\cite{ESM}.

An interesting alternative perspective is apparent if figure~\ref{fig:Rays} is viewed, not as rays diverging from a source, but as rays converging to a focal point.
It is observed that the rays converge to the focal point around the inclusion.
One can envisage several applications where such an effect may be useful.
For example, image distortion from the mirror mounts in telescopes could be reduced by cloaking the mounts.
In addition, apparatus and mounting structures on microwave receivers could be cloaked to improve the quality of the signal.
One could also conceive of cloaking mounting points and the surrounding structures in laser cutting machines to protect them from accidental damage.

\subsubsection{Negative refraction}
It is evident from figure~\ref{fig:Rays} that whilst the rays are continuous, they are not necessarily differentiable.
In particular, at the interface between the cloak and ambient medium, and at the internal interfaces of the cloak, refraction occurs characterised by a discontinuity of the spatial derivatives of the rays.
Of particular interest are the regions on the outer boundary of the cloak where negative refraction occurs.

Consider figure~\ref{fig:Rays-Side}.
Negative refraction occurs on the right hand interface between the cloak and the ambient medium.
A ray exiting the right hand side of the cloak with gradient $M$ at point $\vec{X}^{(0)} = \vec{x}^{(0)}$ may be described by the equation $X_2^{(s)} - X_2^{(0)} = M(X_1^{(s)} - X_1^{(0)})$ in the ambient medium, where $\vec{X}^{(s)}$ is the position of the source.
The behaviour of the ray at the interface is entirely characterised by the position of the source relative to the interface and its initial gradient.
Therefore, the following analysis may be restricted to the right hand side of the cloak.
On the interior of the right hand side of the cloak, the ray exiting the cloak at $\vec{x}^{(0)}$ is characterised by
\[
x_2 = x_1\left(M + \frac{\alpha_1(x_2^{(0)} - Mx_{1}^{(0)})}{x_1-\alpha_2}\right).
\]
The gradient of the ray as it approaches the exterior boundary from the interior of the cloak is then
\[
m^* = \lim_{x_1\to (a+w)^-} \frac{\dd x_2}{\dd x_1} =
\frac{M (a+w-\epsilon)(a+w) - x_2^{(0)}(a+\epsilon)}{w(a+w)}.
\]
Thus, the gradient is discontinuous at the exterior interface.
For negative refraction it is required that $m^*M < 0$, which leads to the following inequalities
\begin{equation}
0 < M < x_2^{(0)}\frac{a+\epsilon}{(a+w)(a+w-\epsilon)}, \qquad\text{or}\qquad
x_2^{(0)}\frac{a+\epsilon}{(a+w)(a+w-\epsilon)} < M < 0.
\label{eq:nr-inequals}
\end{equation}
Assuming the source lies on the line $X_2=0$ as in figure~\ref{fig:Rays-Side} the inequalities reduce to the single inequality
\[
X_{1}^{(s)} < -\frac{(a+w)(w-2\epsilon)}{a+\epsilon},
\]
which is satisfied for all sources outside the cloak $X_{1}^{(s)} < - (a+w)$, if $w< a+3\epsilon$.
Thus, for a sufficiently thin cloak and a cylindrical source placed along $X_2=0$ and any distance from the cloak, negative refraction is expected on the opposite side of cloak.

Alternatively, for a source located along the line $X_1=0$ the inequalities~\eqref{eq:nr-inequals} become
\begin{equation}
0 < X_2^{(s)} < \frac{(a+w)(w-2\epsilon)}{(a+w-\epsilon)}, \qquad\text{or}\qquad
-\frac{(a+w)(w-2\epsilon)}{(a+w-\epsilon)}< X_2^{(s)} < 0,
\end{equation}
where the fact that $|x_2^{(0)}| < (a+w)$ has been used.
Since $a,w > 0$, and $0<\epsilon \ll 1$, the above inequalities are never satisfied.
Hence, the lack of negative refraction on the horizontal interfaces in figure~\ref{fig:Rays-Side}.
Similar arguments may be used to consider other regions where negative refraction may, or may not occur.
It is observed that negative refraction always occurs at the interface between the four regions of the cloak, where the material properties (equivalently the transformation) are not smooth.

\subsubsection{Scattering measure}
\label{sec:error-metric}
It is desirable to have some quantifiable measure of the quality of the cloak with respect to shielding, rather than relying on visual observations.
However, it is not obvious what ``quality'' means with respect to a cloak, given that there are essentially three fields involved, i.e. the ideal field in the absence of both cloak and inclusion, the uncloaked field with an inclusion present but without a cloak, and the cloaked field with both the inclusion and cloak.
Previous experimental works~\cite{Stenger2012} have used an $L_2$ norm computed directly from the measured fields to place a numerical value on the quality of the cloak.
It is in this spirit that the following ``\emph{scattering measure}'' is formally introduced as a tool to quantify the cloaking effect
\begin{equation}
\mathcal{E} (u_1, u_2,\mathcal{R})
=\left(\int\limits_{\mathcal{R}} |u_1(\vec{x}) - u_2(\vec{x})|^2\;\dd \vec{x}\right)
\left(\int\limits_{\mathcal{R}} |u_2(\vec{x})|^2\;\dd \vec{x}\right)^{-1},
\end{equation}
where $\mathcal{R}\subset\mathbb{R}^2$ is some region outside the cloak, and $u_1(\vec{x})$ and $u_2(\vec{x})$ are any two fields.
In the present paper, the quantities $\mathcal{E}(u_u,u_0,\mathcal{R})$ and $\mathcal{E}(u_c,u_0,\mathcal{R})$ are presented for a series of illustrative simulations.
The field $u_0(\vec{x}) = i\Hankel_0^{(1)}(\omega\sqrt{\rho/\mu}|\vec{x}-\vec{x}_0|)/4$ is the Green's function for the unperturbed problem and represents the ``ideal'' field, $u_u(\vec{x})$ and $u_c(\vec{x})$ are the uncloaked and cloaked fields respectively.
Thus, perfect cloaking corresponds to a vanishing $\mathcal{E}$.
Along with the raw scattering measures an additional quantity, $Q = |\mathcal{E}(u_u,u_0,\mathcal{R})-\mathcal{E}(u_c,u_0,\mathcal{R})|/\mathcal{E}(u_u,u_0,\mathcal{R})$ characterising the relative reduction of the scattering measure by the introduction of a cloak is also presented.
It should be emphasised that this is only one of a number of possible measures of quality.

\paragraph{Choice of $\mathcal{R}$.}

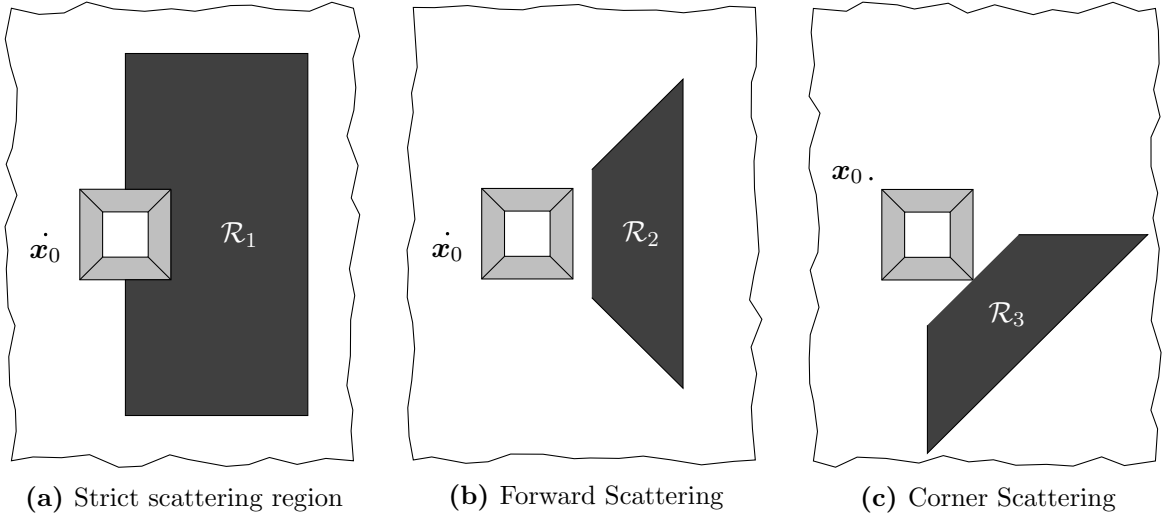
\begin{figure}[htb]
\centering
\begin{subfigure}[c]{0.3\textwidth}
\begin{tikzpicture}[scale=0.3]
\draw[] decorate [decoration={random steps}] {(-5,-10) rectangle (10,10)};
\draw[fill=lightgray] (-2,-2) rectangle (2,2);
\draw[fill=white] (-1,-1) rectangle (1,1);
\draw (1,1) -- (2,2);
\draw (-1,1) -- (-2,2);
\draw (-1,-1) -- (-2,-2);
\draw (1,-1) -- (2,-2);
\draw[fill=darkgray] (2,2) -- (0,2) -- (0,8) -- (8,8) -- (8,-8) -- (0,-8) -- (0,-2) -- (2,-2) -- (2,2);
\node at (5,0) {$\color{white}\displaystyle \mathcal{R}_{1}$};
\node[below] at (-3.5,0) {$\displaystyle \vec{x}_0$};
\draw[fill=black] (-3.5,0) circle (0.05);
\end{tikzpicture}
\caption{\label{fig:R1}Strict scattering region}
\end{subfigure}
\quad
\begin{subfigure}[c]{0.3\textwidth}
\begin{tikzpicture}[scale=0.3]
\draw[] decorate [decoration={random steps}] {(-5,-10) rectangle (10,10)};
\draw[fill=lightgray] (-2,-2) rectangle (2,2);
\draw[fill=white] (-1,-1) rectangle (1,1);
\draw (1,1) -- (2,2);
\draw (-1,1) -- (-2,2);
\draw (-1,-1) -- (-2,-2);
\draw (1,-1) -- (2,-2);
\draw[fill=darkgray] (2.82842712475, 2.82842712475) -- (2.82842712475+4, 2.82842712475+4) -- (2.82842712475+4,-2.82842712475-4) -- (2.82842712475,-2.82842712475);
\node at (5,0) {$\color{white}\displaystyle \mathcal{R}_{2}$};
\node[below] at (-3.5,0) {$\displaystyle \vec{x}_0$};
\draw[fill=black] (-3.5,0) circle (0.05);
\end{tikzpicture}
\caption{\label{fig:R2}Forward Scattering}
\end{subfigure}
\quad
\begin{subfigure}[c]{0.3\textwidth}
\begin{tikzpicture}[scale=0.3]
\draw[] decorate [decoration={random steps}] {(-5,-10) rectangle (10,10)};
\draw[fill=lightgray] (-2,-2) rectangle (2,2);
\draw[fill=white] (-1,-1) rectangle (1,1);
\draw (1,1) -- (2,2);
\draw (-1,1) -- (-2,2);
\draw (-1,-1) -- (-2,-2);
\draw (1,-1) -- (2,-2);
\begin{scope}[rotate=-45]
\draw[fill=darkgray] (2.82842712475, 2.82842712475) -- (2.82842712475+4, 2.82842712475+4) -- (2.82842712475+4,-2.82842712475-4) -- (2.82842712475,-2.82842712475);
\draw[fill=black] (-3.5,0) circle (0.05);
\end{scope}
\node at (3.5,-3.5) {$\color{white}\displaystyle \mathcal{R}_{3}$};
\node[below] at (-3.5,3.5) {$\displaystyle \vec{x}_0$};
\end{tikzpicture}
\caption{\label{fig:R3}Corner Scattering}
\end{subfigure}
\caption{\label{fig:Int-Regs}
The three regions used for computation of the scattering measure.}
\end{figure}

For the purpose of illustration three different regions of integration are considered as shown in figure~\ref{fig:Int-Regs}.
The three regions used were chosen as follows: (a) $\mathcal{R}_1$ is the most strict region used taking into account significant near field effects and a wide range of scattering angles.
However, it is unlikely that this region would be measurable in practice.
(b) The forward scattering region ($\mathcal{R}_2$) is relevant if the scattered field is measurable over a wide range of forward scattering angles.
(c) The corner scattering region ($\mathcal{R}_3$) is employed for sources located along the diagonal of the square inclusion.
It is emphasised that $\|\mathcal{R}_1\|\neq\|\mathcal{R}_2\|=\|\mathcal{R}_3\|$.

In the following section the scattering measures will be presented for a series of illustrative simulations.

\subsection{Illustrative simulations}

\begin{table}[htb]
\centering
\begin{tabular}{@{\qquad}c@{\qquad}c@{\qquad}c@{\qquad}c@{\qquad}c@{\qquad}}
\toprule
\multicolumn{2}{c}{Source} &
\multicolumn{2}{c}{Scattering Measure $\mathcal{E}$}\\
Position & Frequency & Uncloaked & Cloaked & $Q$\\
\midrule
\multicolumn{5}{c}{\underline{Scattering region $\mathcal{R}_1$}}\\
$[-3,0]^\mathrm{T}$ & $5$ & $0.1529$ & $4.351\times10^{-4}$ & 0.9972 \\
$[-3,0]^\mathrm{T}$ & $10$ & $0.1455$ & $4.514\times10^{-4}$ & 0.9969\\
$[-3,3]^\mathrm{T}/\sqrt{2}$ & $5$ & $0.2002$ & $3.941\times10^{-4}$ & 0.9980 \\
$[-3,3]^\mathrm{T}/\sqrt{2}$ & $10$ & $0.3286$ & $4.068\times10^{-4}$ & 0.9988 \\
\midrule
\multicolumn{5}{c}{\underline{Scattering region $\mathcal{R}_2$}}\\
$[-3,0]^\mathrm{T}$ & $5$ & $0.3224$ & $3.664\times10^{-4}$ & 0.9989 \\
$[-3,0]^\mathrm{T}$ & $10$ & $0.3093$ & $1.167\times10^{-3}$ & 0.9962\\
\midrule
\multicolumn{5}{c}{\underline{Scattering region $\mathcal{R}_3$}}\\
$[-3,3]^\mathrm{T}/\sqrt{2}$ & $5$ & $0.2988$ & $3.654\times10^{-4}$ & 0.9988 \\
$[-3,3]^\mathrm{T}/\sqrt{2}$ & $10$ & $0.2988$ & $7.803\times10^{-4}$ & 0.9974 \\
\bottomrule
\end{tabular}
\caption{\label{tab:error-measure-continuum}
The scattering measures corresponding to the simulations shown in figures~\ref{fig:Cont-Cloak-5} and~\ref{fig:Cont-Cloak-10}.}
\end{table}

\begin{figure}[htb]
\centering
\begin{subfigure}[c]{0.35\textwidth}
\includegraphics[width=\textwidth]{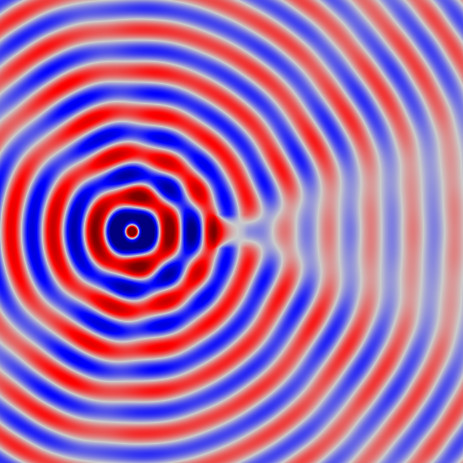}
\caption{\label{fig:C-S-I-U-5}
Uncloaked, $\vec{x}_{0} = [-3,0]^\mathrm{T}$}
\end{subfigure}
\qquad
\begin{subfigure}[c]{0.35\textwidth}
\includegraphics[width=\textwidth]{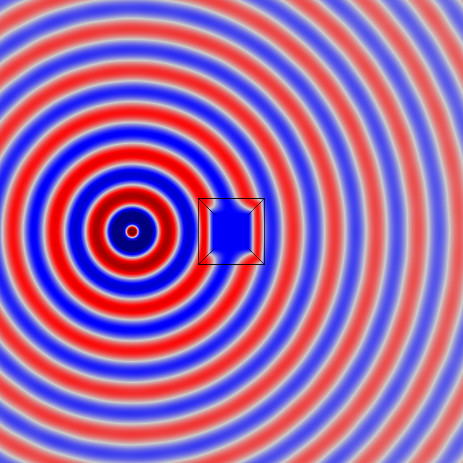}
\caption{\label{fig:C-S-I-C-5}
Cloaked, $\vec{x}_{0} = [-3,0]^\mathrm{T}$}
\end{subfigure}
\qquad
\begin{subfigure}[c]{0.35\textwidth}
\includegraphics[width=\textwidth]{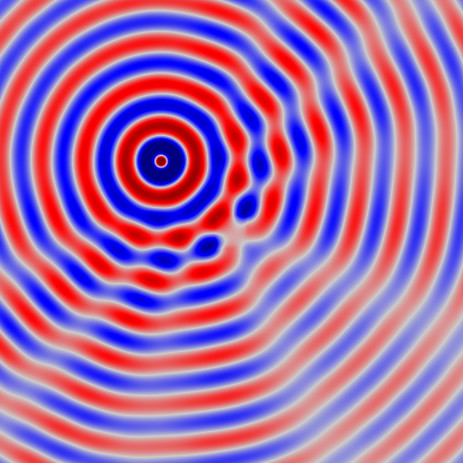}
\caption{\label{fig:C-C-I-U-5}
Uncloaked, $\vec{x}_{0} = [-3,3]^\mathrm{T}/\sqrt{2}$}
\end{subfigure}
\qquad
\begin{subfigure}[c]{0.35\textwidth}
\includegraphics[width=\textwidth]{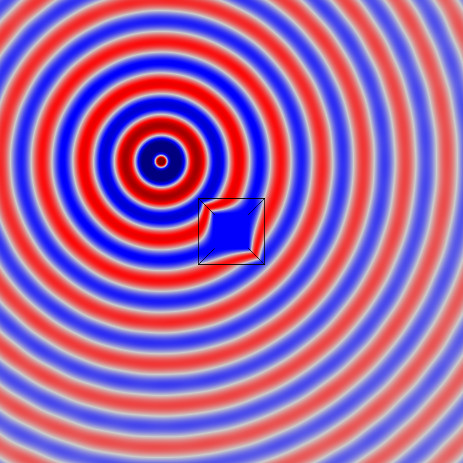}
\caption{\label{fig:C-C-I-C-5}
Cloaked, $\vec{x}_{0} = [-3,3]^\mathrm{T}/\sqrt{2}$}
\end{subfigure}
\caption{\label{fig:Cont-Cloak-5}
Plots of the field $u$ for the uncloaked and cloaked square inclusion, where the angular frequency of excitation is $\omega=5$.
The position of the source is indicated under the relevant plot.
The colour scale is linear from blue (minimum) through white (zero) to red (maximum).}
\end{figure}

\begin{figure}[htb]
\centering
\begin{subfigure}[c]{0.35\textwidth}
\includegraphics[width=\textwidth]{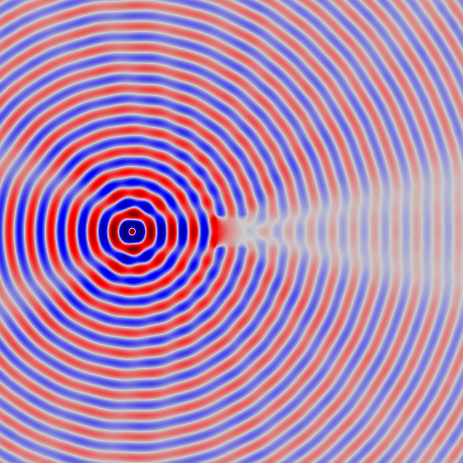}
\caption{\label{fig:C-S-I-U-10}
Uncloaked, $\vec{x}_{0} = [-3,0]^\mathrm{T}$}
\end{subfigure}
\qquad
\begin{subfigure}[c]{0.35\textwidth}
\includegraphics[width=\textwidth]{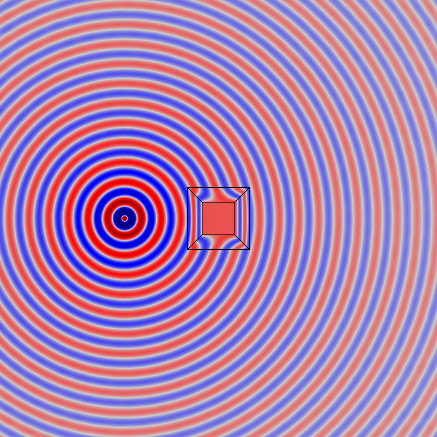}
\caption{\label{fig:C-S-I-C-10}
Cloaked, $\vec{x}_{0} = [-3,0]^\mathrm{T}$}
\end{subfigure}
\qquad
\begin{subfigure}[c]{0.35\textwidth}
\includegraphics[width=\textwidth]{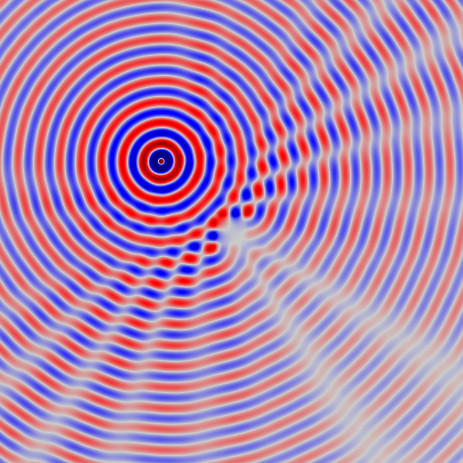}
\caption{\label{fig:C-C-I-U-10}
Uncloaked, $\vec{x}_{0} = [-3,3]^\mathrm{T}/\sqrt{2}$}
\end{subfigure}
\qquad
\begin{subfigure}[c]{0.35\textwidth}
\includegraphics[width=\textwidth]{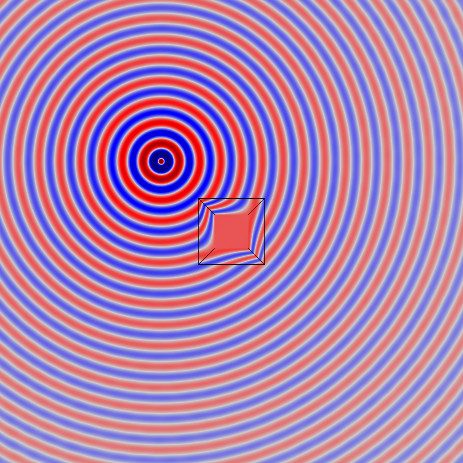}
\caption{\label{fig:C-C-I-C-10}
Cloaked, $\vec{x}_{0} = [-3,3]^\mathrm{T}/\sqrt{2}$}
\end{subfigure}
\caption{\label{fig:Cont-Cloak-10}
Plots of the field $u$ for the uncloaked and cloaked square inclusion where the angular frequency of excitation is $\omega=10$.
The position of the source is indicated under the relevant plot and the inclusion is located at the centre of the image in all cases.
The colour scale is linear from blue (minimum) through white (zero) to red (maximum).}
\end{figure}

\begin{figure}[htb]
\centering
\begin{subfigure}[c]{0.45\textwidth}
\includegraphics[width=\linewidth]{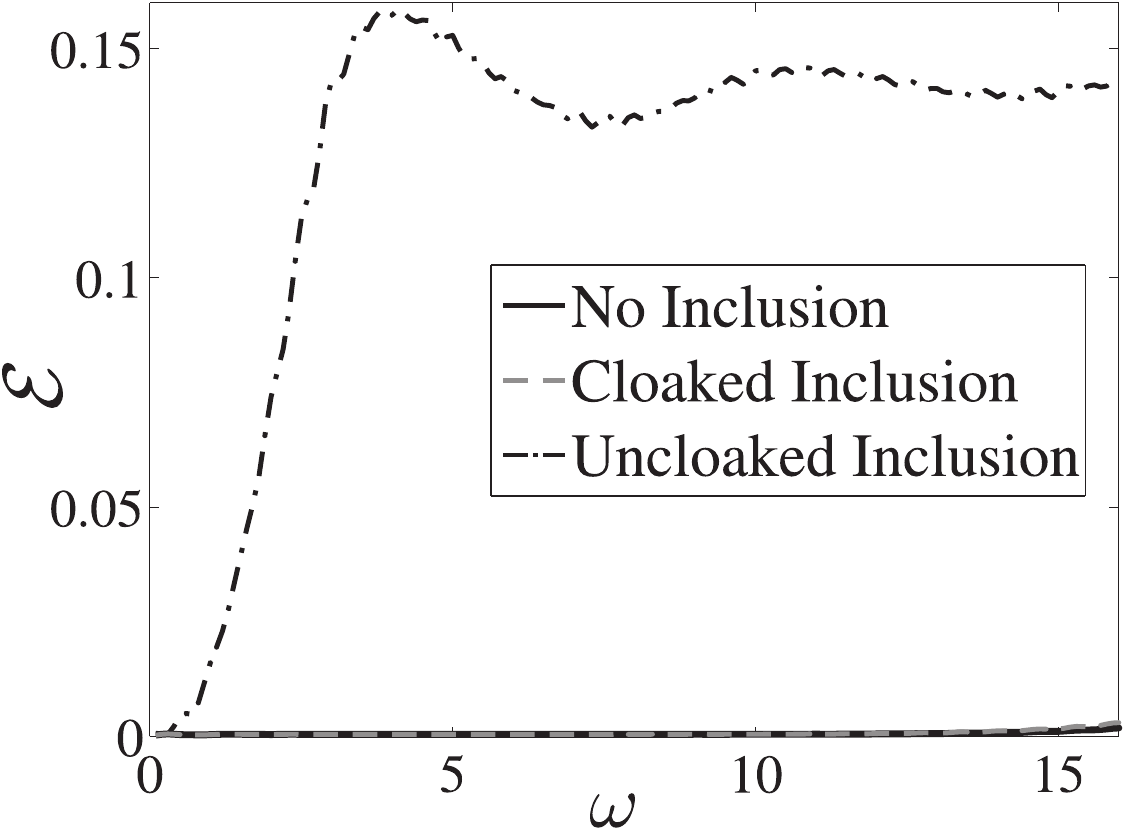}
\caption{\label{fig:error-comp}}
\end{subfigure}
\qquad
\begin{subfigure}[c]{0.45\textwidth}
\includegraphics[width=\linewidth]{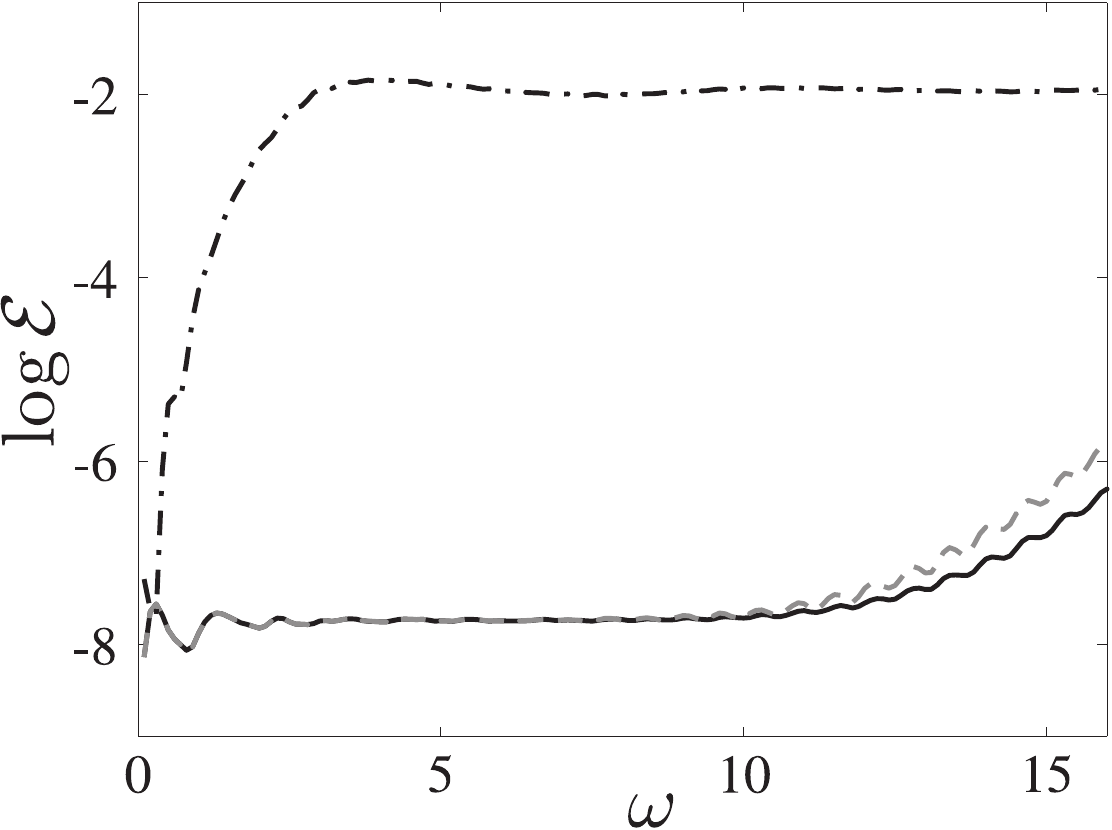}
\caption{\label{fig:log-error-comp}}
\end{subfigure}
\caption{\label{fig:error-comp}
(a) The figure shows the the scattering measure plotted against angular frequency.
(b) The figure shows the log of the scattering measure plotted against angular frequency.
The solid line corresponds to the continuum in the absence of both an inclusion and cloak.
The dashed line represents the cloaked inclusion and the dash-dot line corresponds to the uncloaked inclusion.
The region $\mathcal{R}_1$ (see figure~\ref{fig:Int-Regs} and the associated text) was used to compute the error measure.
}
\end{figure}

A series of illustrative simulations were created using the finite element software COMSOL Multiphysics\textsuperscript{\textregistered}.
For the purposes of these computations, the following non-dimensional parameter values\footnote{Throughout this paper, all numerical parameters are normalised such $\mu=\varrho=1$ unless otherwise stated.} were chosen: $a=0.5$, $w=0.5$, $\mu=\varrho=1$, $\mu_0=0.1$, $\varrho_0=0,\epsilon=1\times10^{-6}$.
Figures~\ref{fig:Cont-Cloak-5} and~\ref{fig:Cont-Cloak-10} show the displacement field $u(\vec{x})$ for a cylindrical source oscillating at $\omega=5$ and $\omega=10$ respectively.
The figures clearly illustrate the efficacy of the square cloak, even at relatively high frequencies.
Table~\ref{tab:error-measure-continuum} shows the corresponding scattering measures as introduced in section~\ref{sec:error-metric}.
It is clear that this square ``\emph{push out}'' cloak is highly effective.
Indeed, for the illustrative simulations presented here, the cloak reduces the scattering measure by not less than $99.62\%$ compared with the uncloaked inclusion.

Figure~\ref{fig:error-comp} shows the scattering measure plotted against non-dimensional angular frequency $\omega$ (with $\mu=\varrho=1$).
The solid curve in figure~\ref{fig:error-comp} corresponds to the continuum, in the absence of both cloak and inclusion.
This curve gives an indication of the numerical error in the simulation induced by, for example, the use of perfectly matched layers and the numerical discretisation.
The dashed curve corresponds to the cloaked inclusion, whilst the dotted curve corresponds to the uncloaked inclusion.
It is observed that the numerical measure of the cloaked inclusion remains close to that of the intact continuum for a large range of frequencies.
Moving to dimensional quantities, suppose the simulation corresponded to a particular polarization of an electric wave travelling through glass at a speed of approximately $2\times10^{8}$ m/s.
The line $\omega=10$ on figure~\ref{fig:error-comp} then corresponds to a frequency of approximately $340$ MHz.

\subsubsection{Boundary considerations}
\begin{figure}[htb]
\centering
\begin{subfigure}[c]{0.35\textwidth}
\includegraphics[width=\textwidth]{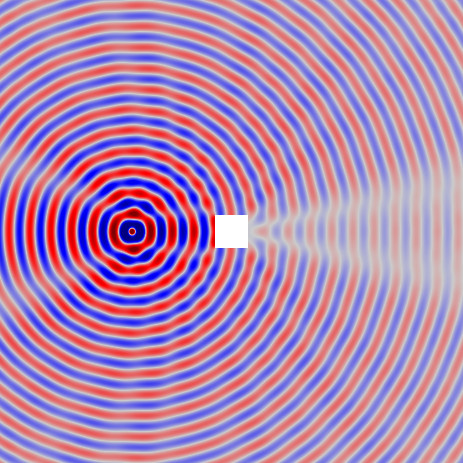}
\caption{\label{fig:C-C-I-U-5}
Uncloaked, Neumann}
\end{subfigure}
\qquad
\begin{subfigure}[c]{0.35\textwidth}
\includegraphics[width=\textwidth]{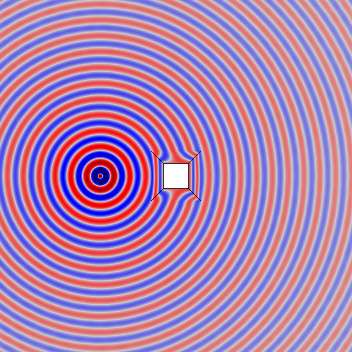}
\caption{\label{fig:C-C-I-C-5}
Cloaked, Neumann}
\end{subfigure}
\begin{subfigure}[c]{0.35\textwidth}
\includegraphics[width=\textwidth]{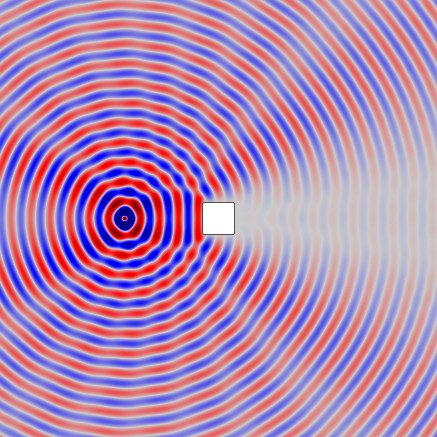}
\caption{\label{fig:C-C-I-U-10}
Uncloaked, Dirichlet}
\end{subfigure}
\qquad
\begin{subfigure}[c]{0.35\textwidth}
\includegraphics[width=\textwidth]{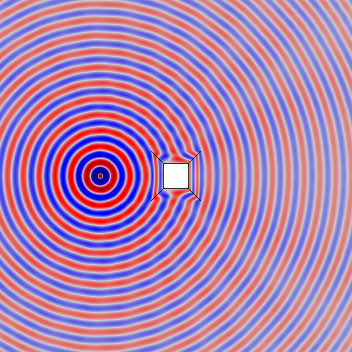}
\caption{\label{fig:C-C-I-C-10}
Cloaked, Dirichlet}
\end{subfigure}
\caption{\label{fig:C-S-DN}
Plots of the field $u$ for the uncloaked and cloaked square inclusion with Neumann boundary conditions in parts (a) and (b), and Dirichlet boundary conditions in parts (c) and (d).
Here the source is located at $\vec{x}=[-3,0]^\mathrm{T}$ and oscillates at $\omega=10$.
The colour scale is linear from blue (minimum) through white (zero) to red (maximum)}
\end{figure}

Whilst cloaking via transformation geometry has been extensively treated in the literature, the sensitivity of the cloaking effect to the boundary conditions is rarely discussed.
The cloak is formed by deforming a small region (a point in the case of the classical radial transformation~\cite{Pendry2006}), into a larger finite region.
If the region is an inclusion, then the natural interface conditions may be determined following the method outlined in section~\ref{sec:interface}.
If the cloaked region is a void or rigid inclusion, however, there is some freedom in choosing the boundary condition, subject to the constraints of the physical problem. 
Figure~\ref{fig:C-S-DN} shows the field $u(\vec{x})$ for a cloaked void, with Neumann (parts (a) and (b)) and Dirichlet (parts (c) (d)) conditions applied to the interior of the cloaked region.
The corresponding scattering measures are shown in table~\ref{tab:error-measure-boundary}.

\begin{table}[htb]
\centering
\begin{tabular}{@{\qquad}c@{\qquad}c@{\qquad}c@{\qquad}c@{\qquad}c@{\qquad}}
\toprule
\multicolumn{2}{c}{Source} &
\multicolumn{2}{c}{Scattering Measure $\mathcal{E}$}\\
Boundary Condition & Frequency & Uncloaked & Cloaked & $Q$\\
\midrule
\multicolumn{5}{c}{\underline{Scattering region $\mathcal{R}_1$}}\\
Neumann & $5$ & $0.1624$ & $4.351\times10^{-4}$ & 0.9973 \\
Neumann & $10$ & $0.1558$ & $4.540\times10^{-4}$ & 0.9971 \\
Dirichlet & $5$ & $0.2931$ & $1.038\times10^{-2}$ & 0.9646 \\
Dirichlet & $10$ & $0.2553$ & $7.875\times10^{-3}$ & 0.9692 \\
\midrule
\multicolumn{5}{c}{\underline{Scattering region $\mathcal{R}_2$}}\\
Neumann & $5$ & $0.3436$ & $3.664\times10^{-4}$ & 0.9989 \\
Neumann & $10$ & $0.3258$ & $1.163\times10^{-3}$ & 0.9964 \\
Dirichlet & $5$ & $0.4864$ & $1.566\times10^{-2}$ & 0.9678 \\
Dirichlet & $10$ & $0.5030$ & $1.673\times10^{-2}$ & 0.9667 \\
\bottomrule
\end{tabular}
\caption{\label{tab:error-measure-boundary}
The scattering measures for a void with Neumann and Dirichlet boundary conditions.
Here the source is located at $[-3,0]^\mathrm{T}$.}
\end{table}

Although the square cloak is effective in both cases, it is clear from both the figures and the table of scattering measures that the type of boundary condition imposed on the cloaked object affects the quality of the cloaking.
Indeed, for a void (Neumann) the cloaking reduces the scattering measure by between $99.7\%$ and $99.9\%$ for both $\omega=5$ and $\omega=10$.
In contrast, cloaking reduces the scattering measure of a rigid inclusion (Dirichlet) by between $96.5\%$ and $96.8\%$ for $\omega=5$ and between $96.9\%$ and $96.7\%$ for $\omega=10$.
The effect of the boundary condition may be interpreted in the following way.
As a result of the transformation, the cloaked object and cloak together behave as if the void is small.
In this sense, the cloaked inclusion represents a singular perturbation of the fundamental solution of the Helmholtz equation.
In the case of a free void with Neumann conditions, the leading order term in the asymptotic expansion is the dipole term, which is of order $\epsilon^2$ and decays like the first derivative of the fundamental solution.
On the other hand, for a fixed void with Dirichlet conditions, the leading order term in the expansion is the monopole term which is of order $\epsilon$ and decays like the fundamental solution.
Thus, the perturbation from the free void is smaller than the perturbation from the fixed void, leading to improved cloaking.

\section{Cloaking path information}
\label{sec:Youngs}

In recent years there has been much interest in experiments to elucidate the fundamental principles of quantum mechanics, and in particular the relationship between measurement and system behaviour.
One basic experiment which with its variants features in many such experimental studies is the classical Young's double slit experiment (see, for example,~\cite{Jacques2007}).
This suggested to us that it would be of interest to consider the interaction of the excellent mechanical cloaking demonstrated earlier with the foundational quantum mechanics experiment.

Thus, a Young's double slit experiment is considered where a monochromatic plane wave is incident on a screen with two apertures.
Due to the superposition of the waves passing through the two apertures, the distinctive double slit interference pattern is produced on an observation screen placed on the opposite side of the apertures to the source.
The result of a simulation of the standard experiment is shown in figure~\ref{fig:Youngs-Intact}, with the diffraction pattern produced on the observation screen (in this case, a vertical line near the right hand edge of figures~\ref{fig:Youngs-Intact}-\ref{fig:Youngs-Cloaked}) shown as curve (a) in figure~\ref{fig:Youngs-Comp}.
Placing an object (inclusion) over one slit, as in figure~\ref{fig:Youngs-Uncloaked}, partially destroys the diffraction pattern.
The corresponding pattern on the observation screen is shown as line (b) in figure~\ref{fig:Youngs-Comp}.
However, coating the object with the square \emph{push out} cloak presented earlier, as shown in figure~\ref{fig:Youngs-Cloaked}, restores the original diffraction pattern almost entirely.
The interference pattern corresponding to the cloaked object is shown as curve (c) in figure~\ref{fig:Youngs-Comp}.

The simulation, shown in the supplementary material, confirms that the excellent cloaking for the inclusion position of figure~\ref{fig:Youngs-Cloaked}, exemplified in figure~\ref{fig:Youngs-Comp}, holds irrespective of the inclusion position.
It has thus been conclusively demonstrated that the cloaking is of sufficient quality to render the interference pattern almost immune to movement in the position of the cloaked obstacle.
In particular movement of the cloaked obstacle, it would seem, does not yield any information about the passage of waves through one slit or the other.
This consideration would be important if one were able to carry out an experiment in which single quantised elements of flexural vibration were in the system at any given instance in time.
The quantum mechanical view would be that, if no path information were available from measurements, the interference fringes behind the double slit should persist.

This proposed quantum experiment raises interesting questions if an appropriate vibration transducer were embedded within the cloak, so that information about vibrations moulded by the cloak were available to experimentalists.
One would assume, in line with the results of say optical experiments of the type referred to in~\cite{Jacques2007}, that any path information gained in this way would be evident in a change in the fringe pattern.
This suggests the interest of a comprehensive quantum mechanical treatment of the interaction between mechanical cloaks and measurement systems.

\section{Cloaking with a lattice}
\label{sec:Lattice-Cloak}
Cloaks designed using transformation optics may have such extreme physical attributes that the requisite materials cannot be physically realised without recourse to metamaterials.
It is with this motivation in mind that the following approximate cloak in the low frequency regime is developed.
The cloak is constructed as an approximation to the continuum square cloak considered earlier, but is realised using a discrete lattice structure, formed from rods and point masses.
The advantage of a discrete structure over a continuous material is that much higher contrasts in material properties are easily realisable using lattices.
The development of an approximate cloaking material using a lattice may allow the practical construction of cloaks.
In the following discussion, it is emphasised that repeated indices are not summed over.

\begin{figure}[htb]
\centering
\includegraphics[width=0.35\textwidth]{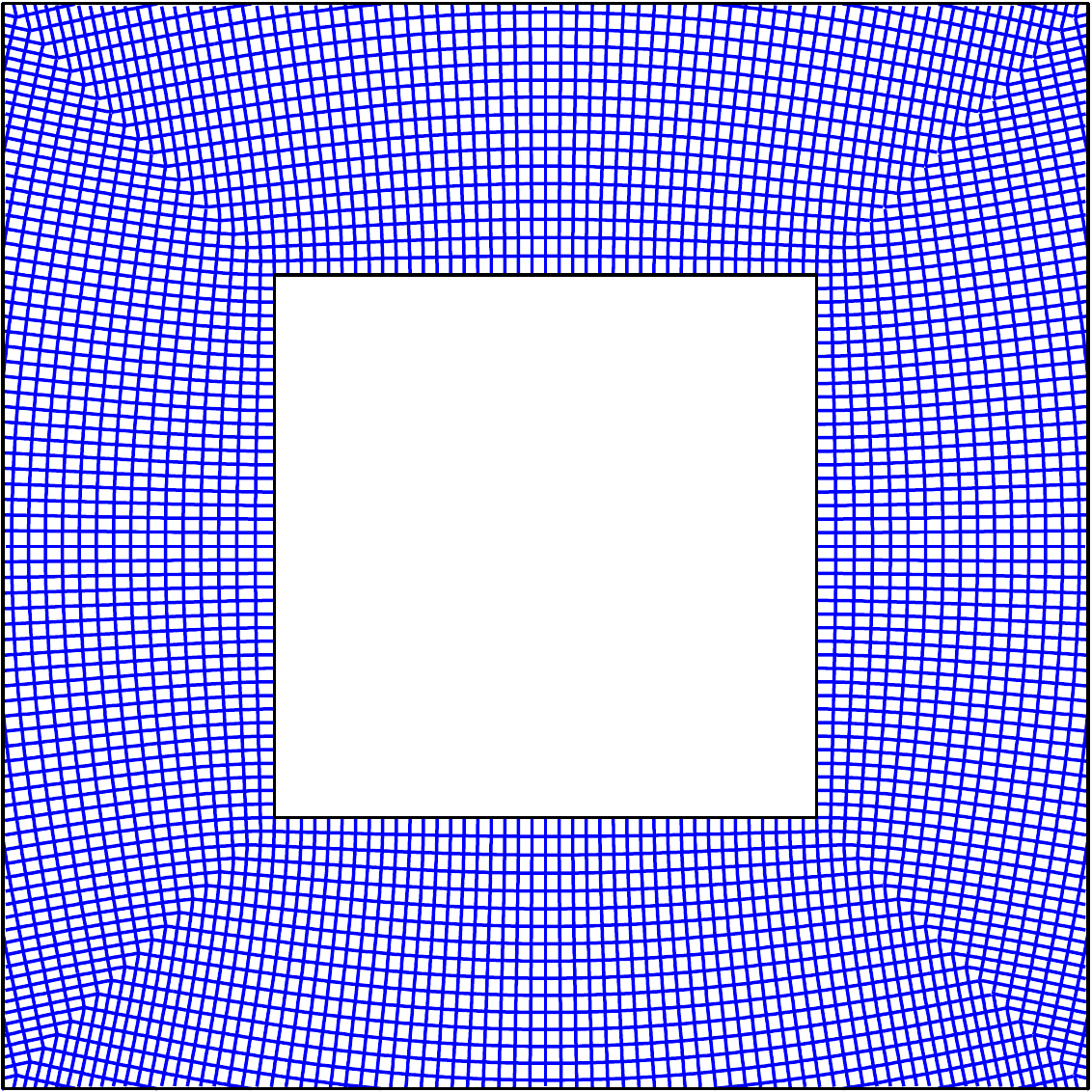}
\caption{\label{fig:principal-lattice}
The lattice formed from the principal directions of the stiffness matrix for the continuum cloak.}
\end{figure}

With reference to the formulae for the Jacaobian of the transformation in section~\ref{subsec:transformation}, the symmetric stiffness matrices $\vec{C}^{(i)} = [\mu/J^{(i)}]{\vec{J}^{(i)}}^\mathrm{T}\vec{J}^{(i)}$ are positive definite.
Therefore, the stiffness matrix admits the following diagonalisation
\begin{equation}
\vec{C}^{(i)} = {\vec{P}^{(i)}}^\mathrm{T}\vec{\Lambda}^{(i)}\vec{P}^{(i)},
\end{equation}
where ${\vec{P}^{(i)}} = [\vec{e}^{(i)}_{1}, \vec{e}^{(i)}_{2}]$ are the matrices with columns consisting of the principal directions (eigenvectors) of  $\vec{C}^{(i)}$, and $\vec{\Lambda}^{(i)} = \mathrm{diag}(\lambda^{(i)}_{1},\lambda^{(i)}_{2})$ is the diagonal matrix of the corresponding ordered [positive] eigenvalues such that $\lambda^{(i)}_{1} > \lambda^{(i)}_{2}$.
The eigenvectors yield the principal lattice vectors of the locally orthogonal lattice with homogenised stiffnesses $\lambda^{(i)}_{j}$ in direction $\vec{e}^{(i)}_{j}$. 
In particular, the lattice nodes lie at the intersection points of the solutions of the following non-linear system of first order differential equations
\begin{equation}
\frac{\dd}{\dd\tau}\vec{x}^{(i)}_j = \vec{e}^{(i)}_j(\vec{x}^{(i)}_j) + \tilde{\vec{x}}^{(i)}_j,\qquad\text{for }i=1,\ldots 4,\text{ and } j=1,2,
\end{equation}
where $\vec{x}^{(i)}_j$ is the position vector along the characteristic defined by $\vec{e}^{(i)}_j$ inside the $i^\text{th}$ side of the cloak, $\tau$ parametrises the curve, and $\tilde{\vec{x}}^{(i)}_j$ is the initial position.
Naturally, this would lead to a lattice with curved links.
However, for a sufficiently refined lattice the curved members may be replaced with linear links.
The lattice links are then the linearisation of the characteristic between two neighbouring nodes on the characteristic.
Figure~\ref{fig:principal-lattice} shows the geometry of the lattice formed from the principal vectors of the stiffness matrix.
Requiring local conservation of flux allows the stiffness of the lattice link parallel to $\vec{e}^{(i)}_{j}$ to be determined as $\ell_{ij}\lambda^{(i)}_{j}$, where $\ell_{ij}$ is the length of the link along $\vec{e}^{(i)}_{j}$ .
The distribution of nodal mass may be determined by evaluating the integral
\[
m(\vec{x_{p}}) = \int\limits_{\mathcal{A}(\vec{x_{p}})} \rho(\vec{x})\, \dd\vec{x}, 
\]
over the unit cell $\mathcal{A}(\vec{x_{p}})$ containing the lattice node at $\vec{x_{p}}$.

In principle, the lattice cloak may be constructed exactly as described above and illustrated in figure~\ref{fig:principal-lattice}.
However, for narrow cloaks where $w/a \ll1$, the locally orthogonal lattice depicted in figure~\ref{fig:principal-lattice} may be approximated by a globally orthogonal regular square lattice.
A regular square lattice is more convenient to implement compared with the non globally orthogonal lattice generated from the eigendecomposition of the stiffness matrix.
Although the geometry of the approximate lattice is regular, it should be emphasised that the stiffness of the links and mass of the nodes vary with position according to the projection of $\vec{A}(\vec{x})$ and $\rho(\vec{x})$ as  described above.

\subsection{Geometry and governing equations for an inclusion cloaked by a lattice}
\label{sec:lattice-geom}
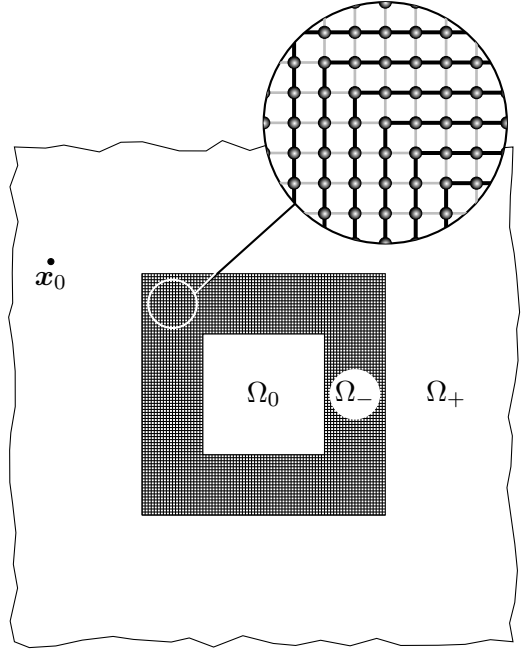
\begin{figure}[htb]
\centering
\begin{tikzpicture}[scale=0.8]
\draw decorate [decoration={random steps}] {(-4.1,-4.1) rectangle (4.1,4.1)};
\draw[fill=white] (-2,-2) rectangle (2,2);

\foreach \i in {-2,-1.95,...,2} {
	\draw (\i,-2) -- (\i,2);
	\draw(-2,\i) -- (2,\i);
}
\draw[fill=white] (-1,-1) rectangle (1,1);
\draw[thick] (-1.15,1.65) -- (2,4.5);
\draw[thick,color=white] (-1.15,1.65) -- (-1.15+0.4*2.85/3.15,1.65+0.4*2.85/3.15);
\draw[thick,fill=none,color=white](-1.5,1.5) circle (0.4);
\draw[thick,fill=white] (2,4.5) circle (2);
\begin{scope}[shift={(2,4.5)}]
	\clip (0,0) circle (2);
	\begin{scope}[rotate=180,color=lightgray,line width=1]
	\draw (0,-10) -- (0,0);
	\draw (0,0) -- (10,0);
	\begin{scope}[shift={(1, -1)}]
		\draw (0,-10) -- (0,0);
		\draw (0,0) -- (10,0);
	\end{scope}
	\begin{scope}[shift={(0.5, -0.5)}]
		\draw (0,-10) -- (0,0);
		\draw (0,0) -- (10,0);
	\end{scope}
	\begin{scope}[shift={(-0.5, 0.5)}]
		\draw (0,-10) -- (0,0);
		\draw (0,0) -- (10,0);
	\end{scope}
	\begin{scope}[shift={(-1, 1)}]
		\draw (0,-10) -- (0,0);
		\draw (0,0) -- (10,0);
	\end{scope}
	\begin{scope}[shift={(-1.5, 1.5)}]
		\draw (0,-10) -- (0,0);
		\draw (0,0) -- (10,0);
	\end{scope}
	\end{scope}
	\draw[line width=1.5] (0,-10) -- (0,0);
	\draw[line width=1.5] (0,0) -- (10,0);
	\begin{scope}[shift={(1, -1)}]
		\draw[line width=1.5] (0,-10) -- (0,0);
		\draw[line width=1.5] (0,0) -- (10,0);
	\end{scope}
	\begin{scope}[shift={(0.5, -0.5)}]
		\draw[line width=1.5] (0,-10) -- (0,0);
		\draw[line width=1.5] (0,0) -- (10,0);
	\end{scope}
	\begin{scope}[shift={(-0.5, 0.5)}]
		\draw[line width=1.5] (0,-10) -- (0,0);
		\draw[line width=1.5] (0,0) -- (10,0);
	\end{scope}
	\begin{scope}[shift={(-1, 1)}]
		\draw[line width=1.5] (0,-10) -- (0,0);
		\draw[line width=1.5] (0,0) -- (10,0);
	\end{scope}
	\begin{scope}[shift={(-1.5, 1.5)}]
		\draw[line width=1.5] (0,-10) -- (0,0);
		\draw[line width=1.5] (0,0) -- (10,0);
	\end{scope}
\foreach \x in {-2,-1.5,-1,-0.5,0, 0.5, 1, 1.5, 2} {
	\foreach \y in  {-2,-1.5,-1,-0.5,0, 0.5, 1, 1.5, 2} {
		\draw[shading=radial, inner color=white, outer color=black] (\x,\y) circle (0.1);
}
}
\end{scope}
\draw[thick,fill=white,color=white](1.5,0) circle (0.4);
\node at (0,0) {$\displaystyle \Omega_0$};
\node at (1.5,0) {$\displaystyle \Omega_-$};
\node at (3,0) {$\displaystyle \Omega_+$};
\node[below] at (-3.5,2.2) {$\displaystyle \vec{x}_0$};
\draw[fill=black] (-3.5,2.2) circle (0.05);

\end{tikzpicture}
\caption{\label{fig:lattice-cloak}
The lattice cloak $\Omega_-$, surrounding the square inclusion $\Omega_0$, embedded in the ambient medium $\Omega_+$.
The thick black lines in the lattice cloak indicate links of high stiffness or conductivity, while the thick grey lines indicate links of low or stiffness conductivity.
}
\end{figure}

Consider a square inclusion $\Omega_0 = \{\vec{x}:|x_1| < a, |x_2|< a\}$, $a>0$, embedded in $\mathbb{R}^2$, surrounded by a cloak $\Omega_- = \{\vec{x}:a<|x_1| < a+w, a<|x_2| < a+w\}\setminus\Omega_0$, where $w > 0$ is the thickness of the cloak.
The cloak consists of a discrete lattice structure with lattice points at $\vec{x} = \ell \vec{p}$, where $\vec{p}\in\mathbb{Z}^2\cap\{\vec{n}: \ell \vec{n}\in\Omega_-\}$.
The lattice is statically anisotropic with links parallel and perpendicular to the boundaries having contrasting material properties, as shown in figure~\ref{fig:lattice-cloak}.

As for the continuum cloak, solutions of the Helmholtz equation are of primary interest.
In particular, the following problem for the field $u(\vec{x})$ is studied
\begin{equation}
[\mu\nabla\cdot(\nabla) + \varrho\omega^2]u(\vec{x}) = -\delta(\vec{x}-\vec{x}_0),
\qquad \vec{x},\vec{x}_0\in\Omega_+,
\label{eq:lattice-helmholtz}
\end{equation}
\begin{equation}
[\mu_{0}\nabla\cdot(\nabla) + \varrho_{0}\omega^2]u(\vec{x}) = 0,
\qquad \vec{x}\in\Omega_0,
\end{equation}
\begin{equation}
\begin{split}
m(\vec{p})\omega^2u(\vec{p}) + \sum_{\vec{q}\in\mathcal{N}(\vec{p})} \ell\eta(\vec{q},\vec{p})\left[u(\vec{p}+\vec{q})- u(\vec{p})\right] = 0,\qquad \text{in }\Omega_-,
\end{split}
\label{eq:lattice-finite-diff}
\end{equation}
where $\vec{e}_i = [\delta_{i1},\delta_{i2}]^\mathrm{T}$,  $\vec{p}\in\mathbb{Z}^2$, and $\mathcal{N}=\{\pm\vec{e}_1,\pm\vec{e}_2\}$ is the set of nearest neighbours.
The stiffness of the lattice links are the restriction of the eigenvalues of the stiffness matrix to the links.
In particular, for the link connecting nodes $\vec{p}$ and $\vec{p}+\vec{q}$, $\eta(\vec{q},\vec{p})$ takes the value $\lambda_{1}^{(i)}|_{[\ell\vec{p},\ell(\vec{p}+\vec{q})]}$ if the vector $\vec{q}$ is parallel to the exterior boundary of the cloak, $\Gamma^{(i)}$, and $\lambda_{2}^{(i)}|_{[\ell\vec{p},\ell(\vec{p}+\vec{q})]}$ otherwise.
The corner regions are matched as illustrated in figure~\ref{fig:lattice-cloak}.
Here, $\lambda_{j}^{(i)}|_{\ell(\vec{q}-\vec{p})}$ indicates the restriction of $\lambda_{j}^{(i)}$ to the line $[\ell\vec{p},\ell(\vec{p}+\vec{q})]$.
The associated interface conditions corresponding to continuity of tractions are
\begin{equation}
\vec{n}\cdot\nabla u(\vec{x}) = \begin{cases}
0 & \text{for } \vec{x} \in \partial\Omega^- \text{ and } \vec{x} \pm \ell\vec{q} \notin \Omega_- \\
\eta(\mp\vec{q},\vec{p}) u(\vec{x}\pm\ell\vec{q})/\mu & \text{for } \vec{x} \in \bigcup_i\Gamma^{(i)} \text{ and }\vec{x} \pm \vec{q} \in \Omega_- \\
\eta(\mp\vec{q},\vec{p}) u(\vec{x}\pm\ell\vec{q})/\mu_0 & \text{for } \vec{x} \in \bigcup_i\gamma^{(i)} \text{ and }\vec{x} \pm \vec{q} \in \Omega_-
\end{cases},
\qquad i=1,\ldots,4,
\label{eq:lattice-boundary-cond}
\end{equation}
and the  Sommerfeld radiation condition at infinity.
The quantity $\eta(\vec{q},\vec{p})$ is the projection of the diagonalised stiffness matrix onto the lattice link connecting lattice points $\vec{p}$ and $\vec{p}+\vec{q}$.

Physically, \eqref{eq:lattice-helmholtz}--\eqref{eq:lattice-boundary-cond} corresponds to the problem of the propagation of time-harmonic waves of angular frequency $\omega$ generated by a point load at $\vec{x}_0$.
The field $u(\vec{x})$ then corresponds to the out-of-plane displacement amplitude and $\varrho$ is the scalar density.
The region $\Omega_-$ consists of an array of nodes of mass $m$, connected by massless rods of length $\ell$
and stiffness according to their orientation.

\subsection{Illustrative lattice simulations}

The approximate lattice cloaks were examined using the finite element software Comsol Multiphysics\textsuperscript{\textregistered}.
PML conditions were imposed on the boundary of the computational domain in order to simulate an infinite domain.
For the purpose of illustration, a square of semi-width $a=0.5$, surrounded by a lattice cloak with $w=0.1$ and links of length $5\times10^{-3}$ was used.
The inclusion is located at the origin of the computational window.

\subsubsection{A basic lattice cloak}

Before proceeding to the illustrative simulations for the regular lattice with heterogeneous distributions of stiffness and mass, it is instructive to consider a simple approximation.
Many cloaks created from transformation optics have the general characteristic of having a high phase speed parallel to the boundary of the cloak, and a phase speed in the direction normal to the boundary (see~\cite{Cummer2007} among others).
Therefore, as an initial approximation, the case of a regular square lattice with a homogeneous, but orthotropic distribution of stiffness and a homogeneous distribution of mass is considered.
Consider the right-hand side of the cloak $\Omega_-^{(1)}$.
For a narrow cloak with $w/a \ll 1$, $x_1\sim a+w$ and hence the density may be approximated by $\rho \sim 1+a/w$.
The greatest contrast in stiffness occurs at $x_2=0$, thus the vertical links are assigned stiffness $\lambda^{(1)}_1(a+w,0)$ and the horizontal links stiffness $\lambda^{(1)}_2(a+w,0)$.
The mass of the nodes is $\ell^2(1+a/w)$.
The material properties of the remaining three sides of the cloak are adjusted accordingly.

\begin{table}[htb]
\centering
\begin{tabular}{@{\qquad}c@{\qquad}c@{\qquad}c@{\qquad}c@{\qquad}c@{\qquad}}
\toprule
\multicolumn{2}{c}{Source} &
\multicolumn{2}{c}{Scattering Measure $\mathcal{E}$}\\
Position & Frequency & Uncloaked & Cloaked & $Q$ \\
\midrule
\multicolumn{5}{c}{\underline{Scattering region $\mathcal{R}_1$}}\\
$[-3,0]^\mathrm{T}$ & $3$ & $0.1430$ & $0.1662$ & 0.1617\\
$[-3,3]^\mathrm{T}/\sqrt{2}$ & $3$ & $0.1113$ & $0.1816$ & 0.6327\\
$[-3,0]^\mathrm{T}$ & $5$ & $0.1529$ & $0.2495$ & 0.6318\\
$[-3,3]^\mathrm{T}/\sqrt{2}$ & $5$ & $0.2002$ & $0.3538$ & 0.7676\\
\midrule
\multicolumn{5}{c}{\underline{Scattering region $\mathcal{R}_2$}}\\
$[-3,0]^\mathrm{T}$ & $3$ & $0.2341$ & 0.3362 & 0.4363\\
$[-3,0]^\mathrm{T}$ & $5$ & $0.3224$ & 0.4671 & 0.4489 \\
\midrule
\multicolumn{5}{c}{\underline{Scattering region $\mathcal{R}_3$}}\\
$[-3,3]^\mathrm{T}/\sqrt{2}$ & $3$ & $0.1578$ & 0.3455 & 1.189\\
$[-3,3]^\mathrm{T}/\sqrt{2}$ & $5$ & $0.2988$ & 0.6011 & 1.012\\
\bottomrule
\end{tabular}
\caption{\label{tab:error-measure-RoT-lattice}
The scattering measures corresponding to the simulations for the \emph{basic lattice model} shown in figures~\ref{fig:Lattice-Cloak-3} and~\ref{fig:Lattice-Cloak-5}.}
\end{table}

Figures~\ref{fig:Lattice-Cloak-3} and~\ref{fig:Lattice-Cloak-5} show the field $u(\vec{x})$ for the uncloaked inclusion (a) and (d), and the inclusion cloaked with this \emph{basic} cloak (b) and (e).
For $\omega=3$ figure~\ref{fig:Lattice-Cloak-3} indicates that the \emph{basic} cloak partially mitigates the shadow cast by the inclusion and acts to reform the cylindrical wave fronts behind the inclusion.
As illustrated by figure~\ref{fig:Lattice-Cloak-5}, this partial cloaking effect deteriorates with increasing frequency.
Indeed, in some cases, the presence of the lattice cloak seems to increase the shadow region.
Table~\ref{tab:error-measure-RoT-lattice} details the values of the scattering measures for the fields illustrated in figure~\ref{fig:Lattice-Cloak-3} and~\ref{fig:Lattice-Cloak-5}.
The scattering measures shown in table~\ref{tab:error-measure-RoT-lattice} suggest that, although visually the basic lattice cloak appears to work reasonably well, this may not be the case.
The fact that the basic lattice cloak increases the scattering measure compared with the uncloaked inclusion further emphasises the need for an objective measure of the quality of cloaks, rather than simply relying on visual observations.

This increase in the scattering measure by the basic lattice cloak motivates the introduction of the following refined model.

\subsubsection{A refined lattice cloak}

\begin{figure}[htb]
\centering
\begin{subfigure}[c]{0.3\textwidth}
\includegraphics[width=\textwidth]{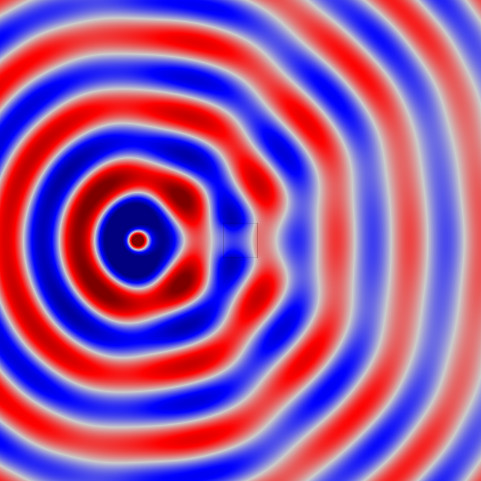}
\caption{\label{fig:RoT-L-S-I-U-3}
Uncloaked}
\end{subfigure}
\quad
\begin{subfigure}[c]{0.3\textwidth}
\includegraphics[width=\textwidth]{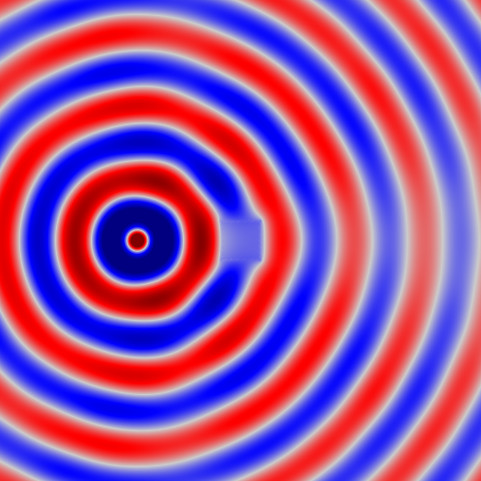}
\caption{\label{fig:RoT-L-S-I-C-3}
Basic cloak}
\end{subfigure}
\quad
\begin{subfigure}[c]{0.3\textwidth}
\includegraphics[width=\textwidth]{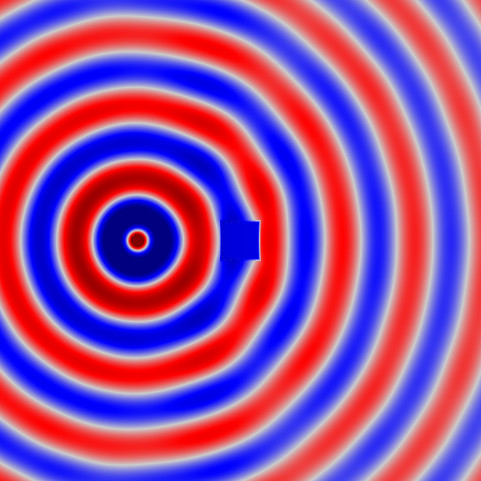}
\caption{\label{fig:L-S-I-C-3}
Refined cloak}
\end{subfigure}
\\
\vspace{2.5ex}
\begin{subfigure}[c]{0.3\textwidth}
\includegraphics[width=\textwidth]{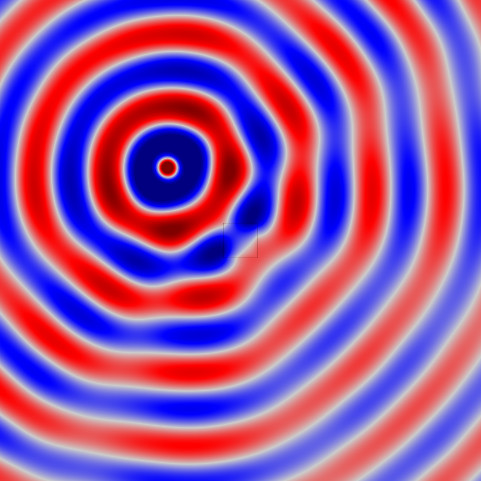}
\caption{\label{fig:RoT-L-C-I-U-3}
Uncloaked}
\end{subfigure}
\quad
\begin{subfigure}[c]{0.3\textwidth}
\includegraphics[width=\textwidth]{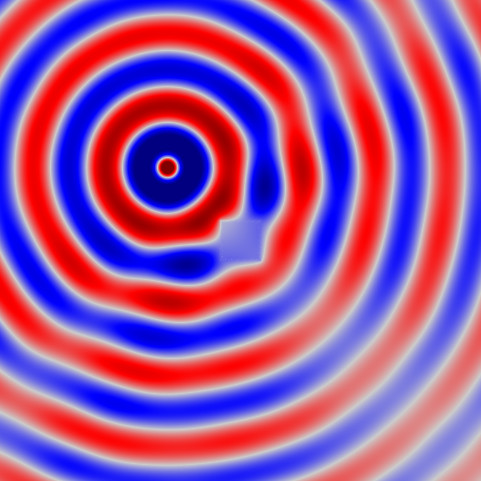}
\caption{\label{fig:RoT-L-C-I-C-3}
Basic cloak}
\end{subfigure}
\quad
\begin{subfigure}[c]{0.3\textwidth}
\includegraphics[width=\textwidth]{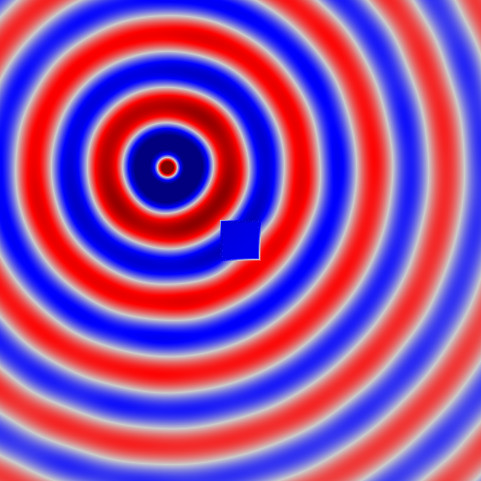}
\caption{\label{fig:L-C-I-C-3}
Refined cloak}
\end{subfigure}
\caption{\label{fig:Lattice-Cloak-3}
Plots of the field $u(\vec{x})$ for a cylindrical wave incident on a square inclusion in the absence of a cloak (parts (a) and (d)), a square inclusion coated with the \emph{basic} lattice (parts (b) and (e)), and an inclusion coating with the refined lattice (parts (c) and (f)).
Here the angular frequency of excitation is $\omega=3$ and the source is located at $\vec{x}_{0} = [-3,0]^\mathrm{T}$ in (a)--(c), and at $\vec{x}_{0} = [-3,3]^\mathrm{T}/\sqrt{2}$ in (e)--(f).
The colour scale is linear from blue (minimum) through white (zero) to red (maximum).}
\end{figure}

\begin{figure}[htb]
\centering
\begin{subfigure}[c]{0.3\textwidth}
\includegraphics[width=\textwidth]{C-S-I-U-5}
\caption{\label{fig:RoT-L-S-I-U-5}
Uncloaked}
\end{subfigure}
\quad
\begin{subfigure}[c]{0.3\textwidth}
\includegraphics[width=\textwidth]{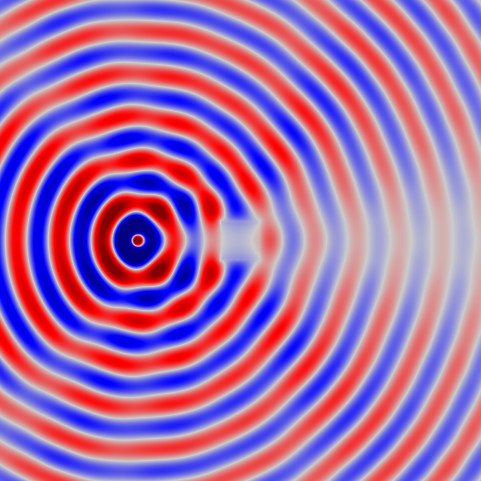}
\caption{\label{fig:RoT-L-S-I-C-5}
Basic cloak}
\end{subfigure}
\quad
\begin{subfigure}[c]{0.3\textwidth}
\includegraphics[width=\textwidth]{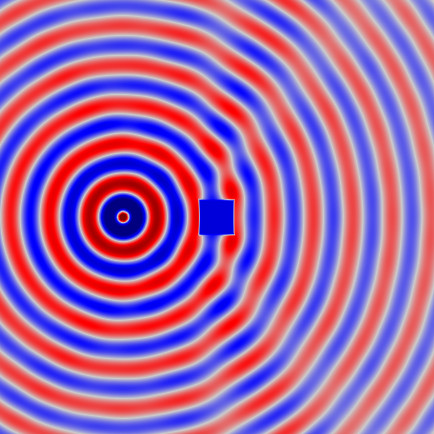}
\caption{\label{fig:L-S-I-C-5}
Refined cloak}
\end{subfigure}
\\
\vspace{2.5ex}
\begin{subfigure}[c]{0.3\textwidth}
\includegraphics[width=\textwidth]{C-C-I-U-5}
\caption{\label{fig:RoT-L-C-I-U-5}
Uncloaked}
\end{subfigure}
\quad
\begin{subfigure}[c]{0.3\textwidth}
\includegraphics[width=\textwidth]{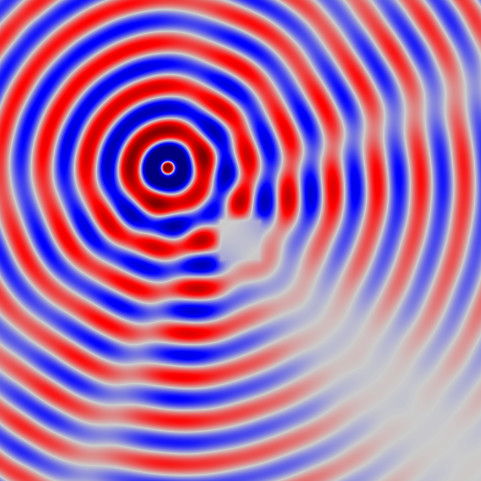}
\caption{\label{fig:RoT-L-C-I-C-5}
Basic cloak}
\end{subfigure}
\quad
\begin{subfigure}[c]{0.3\textwidth}
\includegraphics[width=\textwidth]{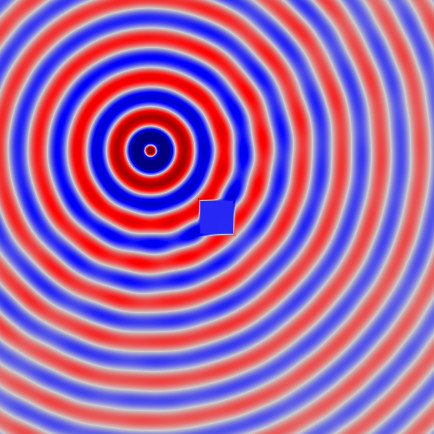}
\caption{\label{fig:L-C-I-C-5}
Refined cloak}
\end{subfigure}
\caption{\label{fig:Lattice-Cloak-5}
Plots of the field $u(\vec{x})$ for a cylindrical wave incident on a square inclusion in the absence of a cloak (parts (a) and (d)), a square inclusion coated with the \emph{basic lattice model} (parts (b) and (e)), and an inclusion coating with the refined lattice (parts (c) and (f)).
Here the angular frequency of excitation is $\omega=10$ and the source is located at $\vec{x}_{0} = [-3,0]^\mathrm{T}$ in (a)--(c), and at $\vec{x}_{0} = [-3,3]^\mathrm{T}/\sqrt{2}$ in (e)--(f).
The colour scale is linear from blue (minimum) through white (zero) to red (maximum).}
\end{figure}

\begin{table}[htb]
\centering
\begin{tabular}{@{\qquad}c@{\qquad}c@{\qquad}c@{\qquad}c@{\qquad}c@{\qquad}}
\toprule
\multicolumn{2}{c}{Source} &
\multicolumn{2}{c}{Scattering Measure $\mathcal{E}$}\\
Position & Frequency & Uncloaked & Cloaked & Q\\
\midrule
\multicolumn{5}{c}{\underline{Scattering region $\mathcal{R}_1$}}\\
$[-3,0]^\mathrm{T}$ & $3$ & 0.1430 & 0.01191 & 0.8929 \\
$[-3,3]^\mathrm{T}/\sqrt{2}$ & $3$ & 0.1113 & $3.385\times10^{-3}$ & 0.9763 \\
$[-3,0]^\mathrm{T}$ & $5$ & $0.1529$ & $0.04324$ & 0.7173 \\
$[-3,3]^\mathrm{T}/\sqrt{2}$ & $5$ & $0.2002$ & $0.03125$ & 0.8438 \\
\midrule
\multicolumn{5}{c}{\underline{Scattering region $\mathcal{R}_2$}}\\
$[-3,0]^\mathrm{T}$ & $3$ & 0.2341 & 0.01150 & 0.9508 \\
$[-3,0]^\mathrm{T}$ & $5$ & $0.3224$ & 0.0172 & 0.9508 \\
\midrule
\multicolumn{5}{c}{\underline{Scattering region $\mathcal{R}_3$}}\\
$[-3,3]^\mathrm{T}/\sqrt{2}$ & 3 & 0.1578 & $5.047\times10^{-3}$ & 0.9680 \\
$[-3,3]^\mathrm{T}/\sqrt{2}$ & $5$ & $0.2988$ & 0.02114 & 0.9292 \\
\bottomrule
\end{tabular}
\caption{\label{tab:error-measure-lattice}
The scattering measures corresponding to the simulations shown for the \emph{refined lattice model} in figures~\ref{fig:Lattice-Cloak-3} and~\ref{fig:Lattice-Cloak-5}.}
\end{table}

Consider now the lattice described in section~\ref{sec:lattice-geom}, i.e. the regular square lattice with inhomogeneous distribution of stiffness and mass.
Figures~\ref{fig:Lattice-Cloak-3} and~\ref{fig:Lattice-Cloak-5} show the field $u(\vec{x})$ for the uncloaked inclusion and the inclusion with a lattice cloaking.
With reference to the simulations for the \emph{basic} cloak (b) and (e) the \emph{refined} lattice cloak (c) and (f), it is observed that the efficiency of the \emph{refined} lattice cloak, whilst not as high as that of the continuum cloak, is much greater than that of the \emph{basic} cloak.
The table of scattering measures for the approximate cloak is shown in table~\ref{tab:error-measure-lattice} and further evidences the effectiveness of the \emph{refined} lattice cloak.
Indeed, for several simulations (in particular those where the scattering measure is taken over the forward or corner scattering regions $\mathcal{R}_1$ and $\mathcal{R}_2$ respectively) the efficiency of the \emph{refined} cloak in reducing the scattering measure approaches that of the continuum cloak.

As expected the effectiveness of the lattice cloaks reduce with increasing frequency.
However, for sufficiently low frequencies the \emph{refined} lattice cloak in particular, works well.

\section{Concluding remarks}
This work represents a comprehensive treatment of a non-singular cloak for a square inclusion.
The significant advantage of this continuous cloak is the straightforward correspondence with a discrete metamaterial lattice structure.
Such a connection may present a method through which a physical cloak may be fabricated.
The material and geometric properties of the discrete cloak are directly linked to the properties of the continuum cloak, and hence, to the properties of the formal map.
The effectiveness of such discrete cloaks, particularly at low frequencies, was demonstrated through numerical simulations and the use of objective scattering measures.

Particular attention was paid to the objective measurement of the quality of the cloaking effect.
The quality of the cloaks were primarily assessed using a scattering measure introduced as an $L_2$ norm of the difference between the cloaked field and the ideal unperturbed field.
A further demonstration of the efficacy of the square \emph{push out} cloak was presented via the classical Young's double slit experiment.
It was shown that the interference pattern on the observation screen was heavily modified when an obstacle was place in front of one of the apertures.
However, if the obstacle was cloaked then the interference pattern remained almost entirely unperturbed.
This numerical experiment presents a further, perhaps more interesting, method through which the quality of particular cloaks may be examined.
Moreover, the experiment raises interesting questions regarding the interaction between cloaking and quantum mechanics.

\paragraph{\footnotesize Acknowledgements:}
{\footnotesize
DJC gratefully acknowledges the support of an EPSRC research studentship \& ABM acknowledges support from EPSRC (EP/H018514/1).
ISJ acknowledges financial support from the EPSRC (EP/H018239/1).
NVM \& RCM acknowledge the financial support of the European Community's Seven Framework Programme under contract number PIAPP-GA-284544-PARM-2.
MB acknowledges the support of the EU FP7 Grant No. PIEF-GA-2011-302357.}

\bibliographystyle{ProcRSocABib}
{\small
\bibliography{Helmholtz-Shield-Refs}
}

\end{document}